\documentclass[]{pasj00}
\Received{2011 February 10}
\Accepted{2011 June 3}
\SetRunningHead{F. Usui et al.}{The AKARI/IRC Mid-infrared Asteroid Survey}

\begin{document}


\title{AcuA: the AKARI/IRC Mid-infrared Asteroid Survey}

\author{
Fumihiko \textsc{Usui}\altaffilmark{1}, 
Daisuke \textsc{Kuroda}\altaffilmark{2}, 
Thomas G. \textsc{M\"{u}ller}\altaffilmark{3}, 
Sunao \textsc{Hasegawa}\altaffilmark{1}, 
Masateru \textsc{Ishiguro}\altaffilmark{4}, \\
Takafumi \textsc{Ootsubo}\altaffilmark{5},  
Daisuke \textsc{Ishihara}\altaffilmark{6},
Hirokazu \textsc{Kataza}\altaffilmark{1}, 
Satoshi \textsc{Takita}\altaffilmark{1}, 
Shinki \textsc{Oyabu}\altaffilmark{6}, \\
Munetaka \textsc{Ueno}\altaffilmark{1}, 
Hideo \textsc{Matsuhara}\altaffilmark{1},
and 
Takashi \textsc{Onaka}\altaffilmark{7} 
}

\altaffiltext{1}{Institute of Space and Astronautical Science, 
Japan Aerospace Exploration Agency, \\
3-1-1 Yoshinodai, Chuo-ku, Sagamihara, Kanagawa 252-5210}
\email{usui@ir.isas.jaxa.jp}

\altaffiltext{2}{Okayama Astrophysical Observatory, 
National Astronomical Observatory, 
3037-5 Honjo, Kamogata, Asakuchi, Okayama 719-0232}

\altaffiltext{3}{Max-Planck-Institut f\"{u}r Extraterrestrische Physik, 
Giessenbachstra\ss e, 85748 Garching, Germany}

\altaffiltext{4}{Department of Physics and Astronomy, 
Seoul National University, 
San 56-1, Shillim-dong Gwanak-gu, Seoul 151-742, South Korea}

\altaffiltext{5}{Astronomical Institute, Tohoku University, 
6-3 Aoba, Aramaki, Aoba-ku, Sendai 980-8578}

\altaffiltext{6}{Graduate School of Science, Nagoya University, 
Furo-cho, Chikusa-ku, Nagoya, 464-8601}


\altaffiltext{7}{Department of Astronomy, Graduate School of Science, 
The University of Tokyo, 7-3-1 Hongo, Bunkyo-ku, Tokyo 113-0033}


\maketitle


\begin{abstract}
We present the results of an unbiased asteroid survey in the mid-infrared
wavelength 
with the Infrared Camera (IRC) onboard the Japanese infrared satellite AKARI.
About 20\% of the point source events recorded in the AKARI All-Sky Survey observations
are not used for the IRC Point Source Catalog (IRC-PSC) in its production
process because of the lack
of multiple detection by position.
Asteroids, which are moving objects on the 
celestial sphere, remain in these ``residual events''. We identify 
asteroids out of the residual events by matching them with the positions of known asteroids. 
For the identified asteroids, we calculate the size and albedo based on the 
Standard Thermal Model. Finally we have a brand-new catalog of asteroids, 
named the Asteroid Catalog Using Akari (AcuA), 
which contains 5,120 objects, about twice as many as the IRAS asteroid 
catalog. The catalog objects comprise 4,953 main belt asteroids, 
58 near Earth asteroids, and 109 Jovian Trojan asteroids. 
The catalog will be publicly available via the Internet. 
\end{abstract}


\section{Introduction}

The physical properties of asteroids are fundamental for the understanding of 
the formation process of our planetary system.
In the present solar system, asteroids are thought to be the primary remnants of 
the original building blocks that formed the planets. They contain a 
record of the initial conditions of 
our solar nebula of 4.6 Gyr ago. The 
composition and size distribution of asteroids in the asteroid belt
provide significant information on their evolution history, although 
they have experienced mutual collisions, mass depletion, 
mixing, and thermal differentiation, which have shaped their present-day 
physical and orbital properties. 

The size and albedo are the basic physical 
properties of the asteroid.  In some cases, 
combining the size and 
the mass that are 
measured precisely by the modern technique (\cite{hilton02}), 
the bulk density of the asteroid can be estimated (\cite{britt02}). 
It is 
a powerful indicator to investigate the macroscopic porosity and the inner 
structure of the asteroid. The total mass and the size distribution of asteroids are 
crucial for the understanding of the history of the solar system (\cite{bottke05}).
The mineralogy and elemental composition of  
asteroids can also be estimated from the albedo (\cite{burbine08}). 

There are several survey catalogs of asteroids: 
the 2MASS Asteroid Catalog (\cite{Sykes00}) compiles 
near-infrared colors of 1,054 asteroids based on the Two Micron All Sky Survey; 
the Subaru Main Belt Asteroid Survey (SMBAS; \cite{Yoshida07}) gives 
the size and color distributions of 1,838 asteroids observed with the Subaru telescope; 
the SDSS Moving Object Catalog (SDSS MOC; \cite{Parker08}) consists of 
multi-color photometry of $\sim$88,000 asteroids from the Sloan Digital 
Sky Survey; the Sub-Kilometer Asteroid Diameter Survey (SKAD; 
\cite{Gladman09}) provides the size distribution of 1,087 asteroids 
based on observations with the 4-m Mayall telescope at Kitt Peak National Observatory.
While these catalogs are based on optical to near-infrared observations, 
the size and albedo of asteroids are decoupled and can be determined solely independently, 
once mid-infrared observations are accomplished
(\cite{lebofsky89, bowell89, harris02}). 

Radiometric technique was first
applied to determine the size and albedo 
of asteroids with ground-based observatories by 
\citet{Al1970} for 4 Vesta, \citet{Allen71} for 1 Ceres, 3 Juno, 4 Vesta,
and \citet{Matson71} for 26 major main-belt asteroids.
A pioneering systematic asteroid survey with a space-borne 
telescope was made by the Infrared Astronomical Satellite (IRAS) launched 
in 1983 (\cite{neugebauer84}). 
IRAS observed more than 96\% of the sky at the mid- and far-infrared 
4 bands (12, 25, 60, and 100 \micron) during the 10 month mission life.
It derived the size and albedo of about 2,200 asteroids (\cite{SIMPS}). 
Another serendipitous survey was carried out by the Midcourse Space Experiment 
(MSX) launched in 1996 (\cite{mill94, Price01}). 
It observed $\sim$ 10\% of the sky 
at 6 bands of 4.29, 4.35, 8.28, 12.13, 14.65, and 21.34 \micron~ 
and about 160 asteroids were identified, for which the
size and albedo were provided (\cite{MIMPS}). Also the Infrared Space Observatory 
(ISO) launched in 1995 (\cite{Kessler96}) made yet-another 
part-of-sky survey and observed several planets, 
satellites, comets, and asteroids at infrared wavelengths (\cite{Mu2002}). 
Despite these extensive past surveys the asteroids for which the size and albedo have been determined are still
only 0.5\% of those with known orbital elements.

AKARI is the first Japanese space mission dedicated to infrared astronomy (\cite{murakami07}).
AKARI is equipped with a 68.5cm cooled telescope, a 170 liter
superfluid liquid Helium (LHe), and 
two sets of two-stage Stirling cycle coolers.
The focal plane instruments consist of the Infrared Camera (IRC)
(\cite{onaka07}) and the Far-Infrared Surveyor (FIS) (\cite{kawada07}), 
each of which covers the spectral range of 2--26\,\micron\, and 50--180\,\micron,
respectively.
AKARI carried out the second
generation infrared all-sky survey after IRAS.  The All-Sky
Survey is one of the main objectives of the AKARI mission
in addition to pointed observations.  It surveyed the whole sky at 6 bands in the mid- to far-infrared 
spectral range with
the solar elongation angle of 90 $\pm$ 1 degree to avoid the
radiation from the Earth and the Sun. The AKARI satellite was launched on
2006 February 21 (UT).  The All-Sky Survey had been continued until the LHe was boiled off on 
2007 August 26. In total, more than 96\% of the sky was observed with more 
than twice (\cite{kataza10}) during the cryogenic 
mission phase. 

In this paper, we present a catalog of the size and geometric 
albedo of asteroids based on the IRC All-Sky Survey data. The IRC All-Sky Survey was carried
out at two bands in the mid-infrared: $S9W$ (6.7--11.6 
\micron) and $L18W$ (13.9--25.6 \micron). 
The IRC All-Sky Survey has advantages over the IRAS survey for detecting asteroids
in the sensitivity and spatial resolution, both of which have been improved by an
order of magnitude.
The 5-$\sigma$ detection limit is 50 and 90\,mJy at the $S9W$ and $L18W$ bands, 
respectively, and
the spatial resolution of the IRC in the 
All-Sky Survey mode was about 10$^{\prime\prime}$ per pixel (\cite{ishihara10}).
Point source detection events 
are extracted and processed in the IRC All-Sky Survey observation data, 
from which the IRC Point Source Catalog 
(IRC-PSC, \cite{ishihara10}) was produced by confirmation
of the source with multiple detection in position. About 20\% of the 
extracted events in the All-Sky Survey data are not used for the IRC-PSC
because of the lack of confirmation detection. 
We identify asteroids out of the excluded events from the 
IRC-PSC. In this process, we search for events whose positions agree with 
those of the asteroids with known orbits.  The asteroid positions are calculated
by the numerical integration of the
orbit.  We do not make an attempt to discover new asteroids in this project, 
whose orbital elements are not archived in the database. 
For each identified object, we calculate 
the size and albedo by using the Standard Thermal Model of 
asteroids (\cite{lebofsky86}). Finally, we obtain an unbiased, 
homogeneous asteroid catalog, which contains 5,120 objects in total, 
twice as many as the IRAS asteroid catalog. This corresponds to about 1\% 
of all the asteroids with known orbital elements.

This paper is organized as follows: In Sect.2, we describe
the data reduction and the creation procedure of 
the asteroid catalog from the IRC All-Sky Survey data. In Sect.3, we describe 
the characteristics of the obtained catalog. 
In Sect.4, we summarize the paper and discuss the future prospect.
Scientific output from this catalog will be discussed at length 
in a forthcoming paper (Usui et al. in preparation). 


\section{Data Processing and Catalog Creation}

\subsection{The AKARI IRC All-Sky Survey}
\label{The IRC All-Sky Survey by AKARI}
The AKARI All-Sky Survey observation
had started on 2006 April 24 as part of the performance verification 
of the instruments prior to the nominal observation, which started on 2006 May 8. 
In the All-Sky Survey observation mode, AKARI always 
points in the direction perpendicular to the Sun-Earth line on
a sun-synchronous polar orbit, and rotates once every orbital 
revolution (see Fig.4 in \cite{murakami07}). The telescope looks 
at the direction opposite to the Earth center to make continuous 
scans on the sky at a rate of 216$^{\prime\prime}$s$^{-1}$. The orbital plane rotates 
around the axis of the Earth at the rate of the orbital motion of 
the Earth, and thus the whole sky can be observed in half a year. 
During the course of the AKARI LHe mission of 18 months, a given area 
of the sky was observed  three or more times on average, depending 
on the ecliptic latitude.  A large number of scan observations were made in 
the ecliptic polar regions, while only two scan observations (overlapping half of the FOV 
in contiguous scans) were possible in  half a year for a given spot on the ecliptic plane. 
In this sense, solar system objects around the ecliptic plane 
have small observation opportunities with AKARI.
In addition to the low visibility, other conditions further limit the observation
opportunities near the ecliptic plane, including the disturbance due to the Moon and the South 
Atlantic Anomaly (SAA), latter of which is a high-density region of charged 
particles (mainly protons) at an altitude of a few hundred km 
above Brazil.  Another complication was introduced to the operation after the first half year, 
which was called the offset survey.
It was an ``aggressive'' operation to swing the scan path to 
complement imperfect scan observations in the first half year, which had been made
in a ``passive'' survey mode.  The first half year survey left
many regions of the sky unobserved as gaps
due to the Moon, the SAA, 
conflicts with pointed observations, and 
data downlink failures to the ground stations. 
To make up observations of  
these regions and increase the completeness of the sky coverage, 
the scan path was shifted from the 
nominal direction to fill the gaps on almost every orbit in the second and
third half years
(Doi et al. in preparation).  For observations of solar system 
objects, the offset survey operation has both positive and negative effects. 
Some objects may lose the observation opportunities completely, 
while others may increase the number of detections drastically. 

\begin{figure}
\begin{center}
\FigureFile(80mm,80mm){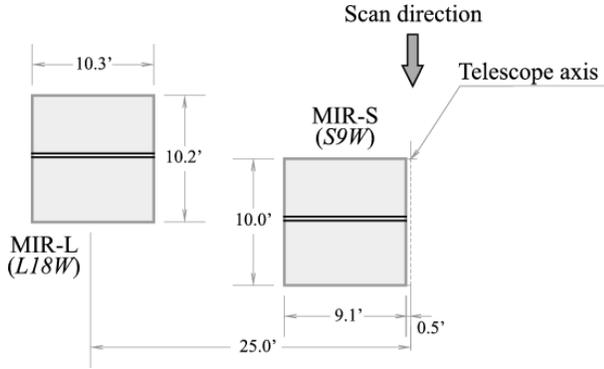}
\end{center}
\caption{
Schematic view of the focal-plane layout of the IRC $S9W$ (MIR-S) and 
$L18W$ (MIR-L) detectors.  Details are given in \citet{murakami07} and \citet{onaka07}. 
The two solid lines in each detector denote the positions of the operating pixel rows 
(the 117th and 125th of the total 256 rows)
for the All-Sky Survey observation mode.  The separation between the two rows is exaggerated 
in this figure and in not in the real scale.
Combining these two rows in the data processing, 
false signals due to cosmic ray hits are removed (milli-seconds confirmation, \citet{ishihara10}).
}
\label{fig:FPI layout}
\end{figure}

Since solar system objects have their orbital motions, the 
detection cannot be confirmed in principle by position
on the celestial sphere. Moreover, $S9W$ and $L18W$ observe 
different sky regions of about 25$^\prime$ apart in the cross-scan direction because of 
the configuration on the 
focal plane (Fig.\ref{fig:FPI layout}) and an object is not observed with the 2 bands in the
same scan orbit. 
Therefore, a single 
event of a point source needs to be examined without stacking 
for detection of asteroids.

\begin{figure}
\begin{center}
\FigureFile(80mm,80mm){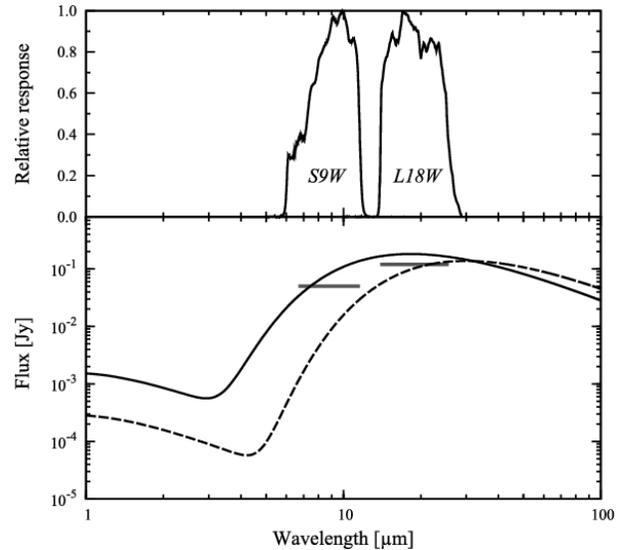}
\end{center}
\caption{Relative spectral response of $S9W$ and $L18W$ from 
http://www.ir.isas.jaxa.jp/ASTRO-F/Observation/RSRF/IRC\_FAD/~. 
(upper panel). As reference, the model spectra of asteroids including 
the reflected sunlight and the thermal emission are shown in the 
lower panel. The solid line indicates 
the model flux of the asteroid with $d=$5km, $p_{\rm v}=0.3$, $R_h=1.56$AU, 
where $d$, $p_{\rm v}$, and $R_h$ are the size (diameter), the geometric albedo, 
and the heliocentric distance, respectively. 
The Standard Thermal Model (Sect. \ref{Thermal model calculation}) 
is used for the calculation. The dashed line indicates that 
with $d=33$km, $p_{\rm v}=0.08$, $R_h=4.6$AU. 
These two asteroids represent a lower limit in the size at the 
corresponding distance in the AKARI survey. 

The horizontal bars in the lower panel are also shown as the detection limits of
$S9W$ and $L18W$.
}
\label{fig:SED}
\end{figure}

Fig.\ref{fig:SED} shows the normalized spectral response function of $S9W$ and $L18W$. 
The calculated model fluxes of asteroids are also shown. 

In the following, we describe how asteroid events are extracted and
identified in the All-Sky Survey observation and how their size and albedo are derived.


\subsection{The Outline of Data Processing}

The outline of the data processing to extract asteroid events 
is summarized in the following (see also Fig.\ref{fig:flow chart}):

\begin{figure}
\begin{center}
\FigureFile(80mm,80mm){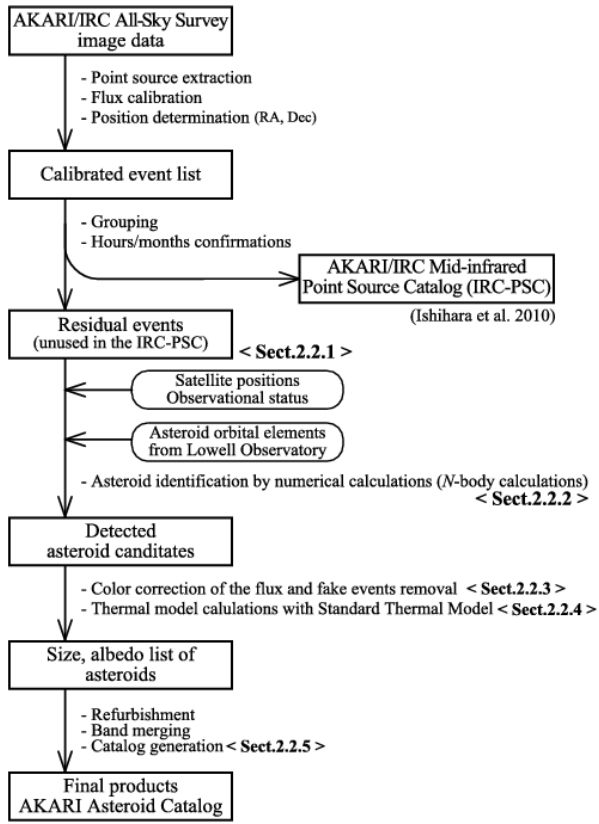}
\end{center}
\caption{Outline of the data processing to create the asteroid catalog.}
\label{fig:flow chart}
\end{figure}

\begin{enumerate}
\item Point sources are extracted by 
the pipeline processing from the IRC All-Sky Survey image data.  
The extracted sources are matched in position with each other
and those detected more than twice are regarded as confirmed sources
and cataloged in the IRC-PSC.  The detected sources not included
in the IRC-PSC are considered to consist of
extended sources, signals due to high energy particles, 
geostationary satellites, and 
solar system objects as asteroids and comets
(Sect.\ref{Event lists for the asteroid catalog}).
Hereafter, individual extracted point sources in the All-Sky Survey
are called  {\it ``events''} and a summary of 
the events is called  {\it ``event list''}.  The physical flux of each event
is derived in the pipeline processing.
\item 
Identification of an event with an asteroid is carried out
based on the predicted positions of the asteroids with known orbital elements 
(Sect.\ref{Asteroid identification}).
\item Color corrections are applied to the fluxes of the events identified as asteroids, 
taking into account the heliocentric distance
of the object.  Events with large errors or those with very small fluxes are discarded from
the list at this stage
 (Sect.\ref{Statistical processing}).
\item The size and albedo of each identified event are calculated based
on the Standard Thermal Model
(Sect.\ref{Thermal model calculation}).
\item Further screening of the sources is carried out and the final catalog is prepared
(Sect.\ref{Final refurbishments and compilations}).
\end{enumerate}


\subsubsection{Event list for asteroid identification}
\label{Event lists for the asteroid catalog}
The present asteroid catalog is a secondary product of the IRC-PSC.  Thus
correction for detector anomalies, image 
reconstruction, point source extraction, pointing 
reconstruction, and  flux calibration are applied in the same manner as in
the IRC-PSC processing (\cite{ishihara10}).  About 25\% ($S9W$) and 
18\% ($L18W$) of the total events are not used for the IRC-PSC and 
they are analyzed in the present process (Table \ref{table:number of events}).

\begin{table*}
\caption{Number of the events for each processing step.}
\label{table:number of events}
\begin{center}
\begin{tabular}{clrr}\hline
& Event & %
\multicolumn{1}{c}{$S9W$}& %
\multicolumn{1}{c}{$L18W$}\\\hline
(a)& All events                      & 4,762,074 & 1,244,249\\
(b)& Events employed in the IRC-PSC  & 3,882,122 &   936,231\\
(c)& Residual events                 &   879,952 &   308,018\\
(d)& Events identified as asteroids  &     6,924 &    13,760\\
(e)& Asteroids in the final catalog  &     2,507 &     5,010\\
(f)& Asteroids detected overall      & \multicolumn{2}{c}{5,120}\\
\hline
\multicolumn{4}{l}{\hbox{\parbox{95mm}{\footnotesize
      \par\noindent
      Notes:(a) {\it ``Event''} indicates an individual detection of a point source in the All-Sky Survey data.
      (b) Events confirmed as a point source by multiple detection at the same celestial position (\cite{ishihara10}).
      (c) Unused events in the IRC-PSC: (c) $=$ (a) $-$ (b).
      (d) Events identified as asteroids by the estimated positions. False identifications 
      are excluded. 
      (e) As is the column name. 
      (f) Asteroids detected with either or both $S9W$ and $L18W$.
      }\hss}}
\end{tabular}
\end{center}
\end{table*}


\subsubsection{Asteroid identification}
\label{Asteroid identification}
Identification with an asteroid is made based on the orbital calculation of the asteroids with 
known orbital elements. $N$-body simulations including 
gravitational perturbations with the Moon, eight planets, Ceres, 
Pallas, Vesta, and Pluto, are employed for the calculation. 
We regard the other asteroids 
as massless particles. The orbital elements of the asteroids are taken 
from the Asteroid Orbital Elements Database (\cite{bowell94}) 
distributed at Lowell 
Observatory\footnote{The data available at ftp://ftp.lowell.edu/pub/elgb/astorb.html}
at the epoch of 2010~04~14.0.  It has 503,681 entries, which consist 
of 233,968 numbered and 269,713 unnumbered asteroids. 
Objects with large uncertainties in the orbital parameters, indicated as 
non-zero integers for the orbit computation in the database,
are excluded.  They include 19 numbered asteroids and 8,759 unnumbered.
The positions of the Sun, planets, Moon, and Pluto are taken from 
DE405 JPL Planetary and Lunar Ephemerides in the J2000.0 equatorial 
coordinates at the NASA Jet Propulsion Laboratory. 
A Runge-Kutta-Nystrom 12(10) method (\cite{dormand87}) is 
used for the time integration with a variable time step.

The asteroid identification process is performed in the following steps:
\begin{enumerate}
\item A 2-body (i.e., the Sun and a given asteroid) 
problem is solved at the epoch of the orbital elements of the 
asteroid to estimate the velocity and acceleration. 
\item Given the observation time of an event detected with 
AKARI, the position of an asteroid is calculated back to that time by the 
$N$-body simulation. The integration time step is initially set as 1 day, and 
varied appropriately later in the following calculations.
The calculated position is 
converted to the J2000.0 astrometric position (i.e., the position is 
revised with the correction for the light-time) since
the positions of the events in the All-Sky Survey are given in the J2000.0 
coordinates.\\
AKARI orbits in a  
sun-synchronous polar orbit at an altitude of 700km. The parallax 
between the geocenter and the spacecraft is not negligible 
particularly for Near Earth Asteroids, which amounts to an order of
a few tens arcsec. 
Thus the apparent position relative to the AKARI spacecraft needs to
be calculated. The spacecraft position is obtained by interpolation of 
the data from the AKARI observational scheduling tool, which has
a sufficient accuracy for the present purpose.
\item The calculated positions are  compared with those of the events
detected in the All-Sky Survey.  If the predicted position of an asteroid is 
located within 2.5$^\prime$ from the position of an event, 
the process goes to the next step.
\item The apparent position of the asteroid is recalculated with a higher
accuracy, taking account of
the correction for the light-time, the gravitational 
deflection of light, the stellar aberration, and the precession and 
nutation of the Earth rotation.  This process takes a large computation time and thus the
calculation is made only for the events tentatively 
associated with an asteroid in the previous step. 
\item The revised position of the asteroid and the position of the corresponding
event are compared again. If the asteroid is located within 7.5$^{\prime\prime}$, 
the position match is regarded as sufficient and the process goes to the next.
\item Then we check the predicted $V$-band magnitude ($M_V$) of the asteroid at 
the observation epoch.  If the predicted $M_V$ is too faint, 
the asteroid should not have been detected with AKARI and 
the identification is regarded as false. 
$M_V$ is calculated by 
using the formulation of \citet{bowell89} with the calculated 
heliocentric distance, ``AKARI-centric'' distance, the absolute 
magnitude ($H$), and the slope parameter ($G$). These $H$-$G$ values are taken
from the dataset of Lowell Observatory as the same file as the orbital 
elements. 
These data mainly originate with the Minor Planet Center.\\
At the same time, the rate of change in the 
right ascension and declination seen from AKARI, the solar elongation 
and moon elongation angle, the phase angle (Sun-asteroid-AKARI angle), 
and the galactic latitude are calculated for later processes. \\
If the object is brighter than $M_V < 23$, the event is concluded 
to be associated as the asteroid.  Otherwise the event is discarded.
\end{enumerate}

\noindent It should be noted that that the 2.5$^\prime$ threshold of the position difference
in step 3 is determined as the maximum value of the correction for
the light-time assuming that a virtual asteroid with the moving speed of
11000 $^\prime$/hr at 0.1 AU from the observer, as:
\begin{eqnarray*}
\frac{11000}{3600} (^\prime/{\rm sec}) \times 0.1({\rm AU}) \times 499.005({\rm sec/AU}) &\sim& 2.5 ^\prime~,
\end{eqnarray*}
and that the 7.5$^{\prime\prime}$ threshold in step 5 is determined as
covering the signal shifted 1 pixel on the detector by chance
where the pixel scale of the detector is
2.3$^{\prime\prime}$,  the FWHM of the point source is 5.5$^{\prime\prime}$
(\cite{kataza10}),
and the position uncertainty including the corrections in step 4
is assumed less than 1$^{\prime\prime}$ .
 

\subsubsection{Color correction and removal of spurious identification}
\label{Statistical processing}
Differences in color between the calibration stars used in the IRC-PSC 
(mainly K- and M-giants, \cite{ishihara10}) and asteroids are not 
negligible because of the wide bandwidth of $S9W$ and $L18W$ 
and the continuum spectra in asteroids that cannot be 
assumed as perfect blackbody or graybodies.  Therefore, we empirically approximate
the color correction factors by a polynomial function
of the heliocentric distance of the object as
\begin{eqnarray}
F_{\rm cc} &=& \frac{F_{\rm raw}}{E_{\rm ccf}} ,\\
\hbox{and}\nonumber\\
{E_{\rm ccf}} &=& a_0 + a_1 R_h + a_2 R_h^{~2} + a_3 R_h^{~3} ~,
\label{eq:ccf}
\end{eqnarray}
\noindent where 
$F_{\rm cc}$, $F_{\rm raw}$, $E_{\rm ccf}$ and $R_h$ are
the color corrected monochromatic flux at 9 or 18\,$\mu$m, the raw in-band flux, 
the color correction factor, and the heliocentric distance, 
respectively. This formula is evaluated using the predicted thermal
flux and the relative spectral response functions of the $S9W$ and $L18W$ bands. 
The predicted thermal flux is calculated by 
assuming that a virtual asteroid with $d=100$km and $p_{\rm v}=0.1$ is located at 
a heliocentric distance between 1.0--6.0AU with a 0.05 AU step, 
where $d$, and $p_{\rm v}$ are the size (diameter), and the geometric albedo, respectively. 
We determine the coefficients $a_0$, $a_1$, $a_2$, and 
$a_3$ as listed in Table \ref{table:color correction factor}. 
The fitting errors of equation 
(\ref{eq:ccf}) to the calculated model flux is 6\% for $S9W$
and 2.5\% for $L18W$ at most. 
The actual value of $1/E_{\rm ccf}$ is in a range
of 1.06 -- 0.80 for $S9W$ and 1.07 -- 0.99 for $L18W$ for the heliocentric
distance of 1--6 AU.

\begin{table}
\caption{The coefficients of the color correction factors.}
\label{table:color correction factor}
\begin{center}
\begin{tabular}{c|llll}
\hline
     & $a_0$   & $a_1$     & $a_2$     & $a_3$\\\hline
$S9W$ & $0.984$ & $-0.068$ & $0.031$  & $-0.0019$\\
$L18W$ & $0.956$ & $-0.024$ & $0.007$  & $-0.0003$\\\hline
\end{tabular}
\end{center}
\end{table}

Up to this stage, the flux level of the event is not taken into account
in the identification procedure.  We discard false identifications
in the following steps based on the flux level.

\begin{itemize}
\item Events with extremely large uncertainties in the flux are 
discarded. Here we set the threshold for the flux uncertainty as 71Jy for $S9W$
and 96Jy for $L18W$.  
These threshold values are determined by a 5$\sigma$ clipping method, 
i.e., the standard deviation ($\sigma$) of 
distribution of the flux uncertainties for all events is
determined and the events of outside 5$\sigma$ values are
discarded.
47 events at $S9W$ and 
101 events at $L18W$ are discarded based on these criteria. 
This step in fact excludes events
affected by the stray light near the Moon efficiently.
\item 
The faintest sources in the IRC-PSC have
fluxes of 0.045 Jy at $S9W$ and 0.06 Jy at $L18W$ (\cite{ishihara10}). 
These values correspond to the signal-to-noise ratio ($S/N$) of 
6 and 3, respectively.  There are a few events that have fluxes fainter than
these values in the event list.  Because of the low $S/N$ of the fluxes, it is
difficult to derive the size and albedo accurately for these objects.
Thus these objects are also excluded from the catalog.

\end{itemize}


\subsubsection{Thermal model calculation}
\label{Thermal model calculation}
Radiometric analysis of the identified events is carried out  with the 
calibrated, color corrected, monochromatic fluxes described in 
Sect.\ref{Statistical processing}. We used a modified version of 
the Standard Thermal Model (STM; \cite{lebofsky86}). 
In the STM, it is assumed that an asteroid is 
a non-rotating, spherical body, and 
the thermal emission from the point on an asteroid's surface is in instantaneous 
equilibrium with the solar flux absorbed at that point. 
Then, the temperature distribution, $T$, on a smooth spherical surface of asteroid 
is simply assumed to be symmetric with respect to the subsolar point as: 
\begin{equation} 
T(\varphi) = \left\{ \begin{array}{lcl}
T_{\rm SS} \cdot \cos^{1/4} \varphi~ ,  & \mbox{for} & \varphi \leq {\pi}/{2}~ , \\
0~ ,                                    & \mbox{for} & \varphi > {\pi}/{2}~ , \\
\end{array} \right. 
\label{eq:temperature}
\end{equation}
where $\varphi$ is the angular distance from the subsolar point. This assumes that 
the temperature on the nightside is treated as zero. 
The subsolar temperature, $T_{\rm SS}$, is determined by equating the energy balance so that 
the absorbed sunlight is instantaneously re-emitted at thermal infrared wavelengths, thus,
\begin{eqnarray}
T_{\rm SS} &=& \left(\frac{(1-A_{b})S_{s}}{\eta \varepsilon \sigma R_{h}^{~2}}\right)^{1/4}~ , 
\label{eq:max temperature}
\end{eqnarray}
where $A_{b}$, $S_{s}$, $\eta$, $\varepsilon$, and $\sigma$ are 
the bond albedo, the incident solar flux, 
the beaming parameter, the infrared emissivity, and the Stefan-Boltzmann constant, respectively. 
It is usually assumed that 
\begin{eqnarray}
A_{b} &=& q p_{\rm v}~ , 
\end{eqnarray}
where $q$, and $p_{\rm v}$ are the phase integral, and the geometric albedo. 
The phase integral $q$ is given (the standard H-G system; \cite{bowell89}) by
\begin{eqnarray}
q = 0.290 + 0.684 G~ ,
\label{eq:phase integral}
\end{eqnarray}
where $G$ is the slope parameter. 

\noindent The scattered light is observed at optical to near-infrared wavelengths. 
The diameter can then be derived from the relation as:
\begin{eqnarray}
d &=& \frac{1329}{\sqrt{p_{\rm v}}}~10^{-H/5} ,
\label{eq:relation of d and H}
\end{eqnarray}
where $d$, and $H$ are the diameter in units of km, and the absolute magnitude, respectively
(see e.g., \citet{Fowler92}). 

In applying the STM model, the parameters $H$ and $G$, which are used in the 
identification process (Sect.\ref{Asteroid identification}), are also employed as an optical flux. 
The infrared emissivity $\varepsilon$ is assumed to be a constant of 0.9 as a standard value 
for the mid-infrared. 
The geometry is determined by 
the heliocentric distance, the AKARI-centric distance, and the phase
angle. 
We assume a thermal infrared phase coefficient of 0.01 mag/deg as 
specified for the STM. For the error calculation we assign 
uncertainties of 0.05 mag for $H$ and 0.02 for $G$. 
While the beaming parameter $\eta$ basically accounts for the 
physical quantities relating to the surface roughness and the thermal 
inertia of the asteroid, it is used just as an empirical parameter, 
particularly in the STM.  

The thermal flux of the model is calculated 
by integrating the Planck function numerically with equation (\ref{eq:temperature}) 
over a spherical asteroid of the diameter $d$ 
under the condition of equation (\ref{eq:relation of d and H}). 
The process is iteratively examined until the model flux converges on the observed value 
by adjusting the variables of $d$ and $p_{\rm v}$. 

In the first analysis we concentrate on 55 selected, well-studied 
main-belt asteroids (\cite{mueller05}), 
whose size, shape, rotational property, and albedo are known from different measurements (occultation, direct imaging, 
flybys, and radiometric techniques based on large thermal datasets) 
as listed in Table \ref{table:55 asteroids} in Appendix \ref{List of 55 asteroids of calibrators}. These samples include
asteroids of the sizes between about 70 and 1000 km and the albedos 
from 0.03 to 0.4. The verification of the STM approach 
for a given AKARI asteroid is examined with this dataset. \citet{lebofsky86} made 
a similar exercise for 1 Ceres and 2 Pallas and derived a 
beaming parameter of $\eta$ = 0.756 to obtain an acceptable match between 
the radiometrically derived size and albedo from 
$N$- and $Q$-band fluxes of ground-based observations 
and the published occultation 
diameters.  
For the AKARI dataset of $S9W$ and $L18W$, 
we adjust the beaming parameter to obtain the
best fit in the size and albedo between the values derived from the AKARI
2-band data and the known values.  The best 
fit is obtained with $\eta$ = 0.87 for $S9W$ and 0.77 for $L18W$. 
We also 
attempt to fit the 2-band data simultaneously with a single $\eta$ for the
objects for which both data are available at the same epoch. 
However the overall 
match becomes significantly worse.  We therefore decided to use 
different values of $\eta$ for each band. 


\subsubsection{Final adjustment and creation of the catalog}
\label{Final refurbishments and compilations}
Thermal model calculations provide  
unreasonable values (either too bright or too dark) for some asteroids. 
They are regarded as false identification.
We set the threshold for albedo as $0.01 < p_{\rm v} < 0.9$ and those outside of the
range are discarded.
The number of the discarded events at this stage is 178 for $S9W$
and 53 for $L18W$, about 1 \% of the total identified events.

To obtain the final product, we take a mean of the 
size and albedo with the weight of the $S/N$ for each object. 
For the IRC All-Sky Survey data, the
$S/N$ is given as a function of the measured flux (see Fig.15 in 
\cite{ishihara10}). For the asteroids, about 68\% of $S9W$ and 74\% 
of $L18W$ events reach the maximum $S/N$ values,  $S/N=15$ for $S9W$ 
and $S/N=18$ for $L18W$.  The corresponding flux is about 0.6Jy at $S9W$
and 1.0Jy at $L18W$. If all the fluxes of an asteroid are above these values, 
the weighted mean is equal to a simple 
arithmetic mean. 

Finally,  a total of 5,120 objects (5,079 numbered, and 41 unnumbered 
asteroids) are included in the catalog of the AKARI Mid-infrared 
Asteroid Survey, named the Asteroid Catalog Using Akari (AcuA). 


\section{Evaluation of the asteroid catalog}
\subsection{Uncertainty of the catalog data}
One of the most major contributions which bring uncertainties 
in the size and albedo is the uncertainty of the observed 
fluxes of the asteroids. It is expressed in terms of the 
$S/N$ of the fluxes of the events in the IRC-PSC. 
As mentioned in Sect.\ref{Final refurbishments and compilations}, 
the $S/N$ have reached a plateau at $S/N=15$ for $S9W$ and 
$S/N=18$ for $L18W$.  
Thus even for the best cases the uncertainties in the fluxes
are 6.7\% and 5.6\%, for $S9W$ and $L18W$, respectively. This is directly result
in uncertainties in the size of 3.3\% and 2.8\% and in the albedo of 6.7\% and 5.6\%. 
It is inherent component in this work. 

The absolute magnitude ($H$) is adopted from the same dataset of 
Lowell Observatory as for the orbital elements
described in Sect.\ref{Asteroid identification}.  The uncertainty 
in $H$ is given as three levels: 0.5, 0.05, and 0.005mag in the dataset. 
We suspect that the visual absolute magnitude $H$ has a large
uncertainty and probably larger than those cataloged in some cases.
Thus we decided to give a constant uncertainty of 0.05 mag
for objects listed with the uncertainties of 0.005mag (963 asteroids) and
0.05 mag (4,157 asteroids) of 
our 5,210 cataloged asteroids rather than to use the original uncertainties
in the dataset. 
This corresponds to a 4.6\% uncertainty in 
albedo and less in size. 
The slope parameter ($G$) is also taken from the dataset of
Lowell Observatory. In our cataloged asteroids, 5,015 
objects are assumed as $G=0.15$ and others are provided 
severally.  The uncertainty of $G$ is assumed as 0.02 
uniformly.  It has a small influence on the derived size
and albedo as expected in
equation (\ref{eq:phase integral}). 

In our catalog, these three parameters, i.e., the observed fluxes, the absolute magnitude $H$, 
and the slope parameter $G$ are considered as the contributed factors for the uncertainties
in the size and albedo. 
From these combination, typical values of uncertainties in size is 4.7\%, and those in albedo is 10.1\%. 
The other components discussed below are not used for the uncertainty
calculation because they are not appropriately quantified in this work.

In this work, we apply the STM 
(Sect.\ref{Thermal model calculation}) to derive the size and albedo. 
It is assumed that
an asteroid is a non-rotating, spherical body at a limit
of zero thermal inertia.  Thus the flux variation
due to rotation of an object is neglected. Detailed 
investigations require further information on the object, 
such as the individual shape model, the direction of 
the spin vector, and so on. Since continuous observations 
with AKARI have at least a 100 minute interval (one orbital 
period of the satellite) inevitably, light curves with 
fine time resolution cannot be obtained. 
Therefore it is difficult to determine the detailed model 
parameters solely by AKARI observations.
It is known
that many asteroids have large amplitude ($\sim$ a few 
tens \%) in the light curves (\cite{Warner09}). This adds
a few to $\sim$ 10\% uncertainties in size, especially for those
with a small number of detections.  
Therefore the uncertainties
in the size and albedo originating from the flux uncertainty
could be larger for those asteroids.

The model parameters in the STM are the emissivity ($\varepsilon$),
the thermal infrared phase coefficient, and the beaming 
parameter ($\eta$). The former two are given as fixed values
in advance. Because of a severe constraint on the solar 
elongation angle, observations with AKARI cannot be made 
with several different phase angles. For this reason, the phase 
coefficient is fixed as 0.01mag/deg in the present analysis (\cite{Matson71}).
The different values are used for the beaming parameter $\eta$ 
for $S9W$ and $L18W$. The different values are chosen to adjust the
derived size and albedo to those reported in previous work.
The failure of any single value of $\eta$ to provide good results
with previous work may stem from the invalid assumptions in the
STM.  The beaming parameter is in fact not a physical quantity,
but rather introduced to account for the observations empirically.
AKARI did not observe an asteroid with the two bands
simultaneously, which could affect the way of the adjustment of
$\eta$ at the two bands.
The uncertainty of $\eta$, a 5\% change 
in $\eta$ leads to about 4\% and 2\% changes in size and 
about 8\% and 5\% changes in albedo at $S9W$ and $L18W$, 
respectively, depending slightly on the albedo of the object. 

The geometry is given by the heliocentric distance, 
the AKARI-centric distance, and the phase angle. These are
dependent on the position accuracy of the IRC-PSC 
(less than $2^{\prime\prime}$, \citet{ishihara10}), 
and their uncertainties are negligible for the 
obtained catalog values.

\subsection{Total number and spatial distribution}
The number of the asteroids identified in the AKARI All-Sky Survey is summarized
in Table \ref{table:number of events}.  The net number of the asteroids 
detected with $S9W$ and $L18W$ in total is 5,120.  
The number of the asteroids
detected at $L18W$ is by about twice larger than that at $S9W$.  
The number of the point sources detected at $S9W$ in the 
IRC-PSC is approximately four times as many as that at $L18W$.  The 
opposite trend can be explained by the different spectral energy 
distribution of the objects:
asteroids have typical effective temperatures around 200K and radiate 
thermal emission with the peak wavelength of $\sim$15 \micron, which can
preferentially be detected at $L18W$ even if the difference in the sensitivity
is taken account (Fig.\ref{fig:SED}).  Stellar sources emit
radiation with the peak wavelength at UV to optical and
thus are detected with higher probabilities at $S9W$.
A significant fraction of asteroids, particularly in the main-belt rather
than the near-Earth, is detected only at $L18W$, but undetected at $S9W$
because of the steep decrease in the thermal radiation in the Wien's
domain.  

\begin{figure}
\begin{center}
\FigureFile(80mm,80mm){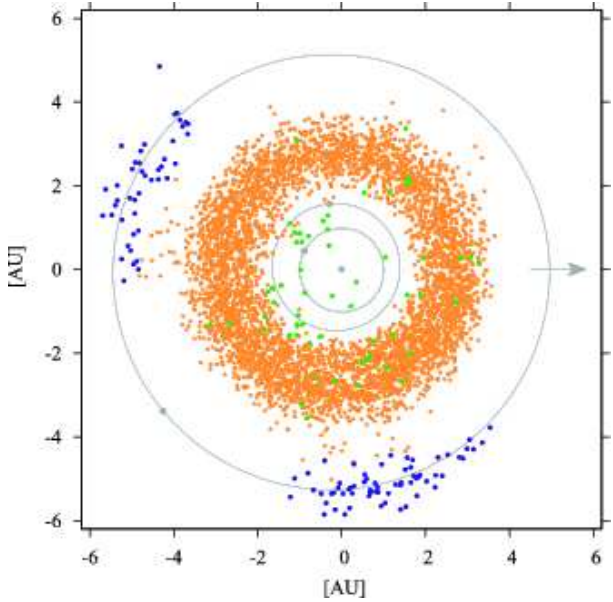}
\end{center}
\caption{Distribution of the identified asteroids projected on the 
ecliptic plane as of 2006 February 22.  The circles indicate the orbits of 
the Earth, Mars, and Jupiter from inside to outside.  
The orange, green, and blue dots indicate the MBAs, the NEAs, and the Trojans, respectively. 
The arrow shows the direction of the vernal equinox.}
\label{fig:plot on ecliptic plane}
\end{figure}

In Fig.\ref{fig:plot on ecliptic plane}, we show the 
distribution of the identified asteroids projected on the ecliptic 
plane (i.e., the face-on view). The near Earth asteroids (NEAs),
 the main-belt asteroids (MBAs), and the Jovian Trojans can be discerned in the plot, 
while Centaurs and Trans-Neptune objects are not detected in our 
survey.  
Fig.\ref{fig:plot on ecliptic plane} displays the location of the 5,120 asteroids at the epoch
of 2006 February 22.  It shows the distribution of asteroids
without any bias and survey gap.

\subsection{Number of detections per asteroid}

\begin{figure}
\begin{center}
\FigureFile(80mm,80mm){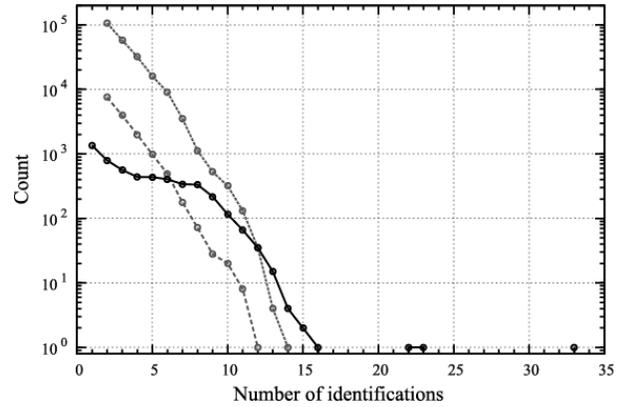}
\end{center}
\caption{Histogram of the number of detections of the asteroids 
identified with the AKARI All-Sky Survey (solid line). The objects with 
extremely large numbers are
137805 (1999 YK5) with 33, 
P/2006 HR30 (Siding Spring) with 23, 
85709 (1998 SG36) with 22, 
and 366 Vincentina (1893 W) with 16 detections. 
The gray dashed and the gray dotted lines show
the number of the events with the sum of $S9W$ and $L18W$,
which are used as input to the IRC-PSC (\cite{kataza10}),
for $|\beta| < 1^\circ$ and $|\beta| < 15^\circ$, respectively,
where $\beta$ is the ecliptic latitude of the source.
}
\label{fig:histogram of NID}
\end{figure}

\begin{figure*}
\begin{center}
{\small
\begin{tabular}{cc}
\hspace{20pt}137805 (1999 YK5): 33 detections & \hspace{20pt}P/2006 HR30 (Siding Spring): 23 detections\\[-0pt]
\FigureFile(60mm,60mm){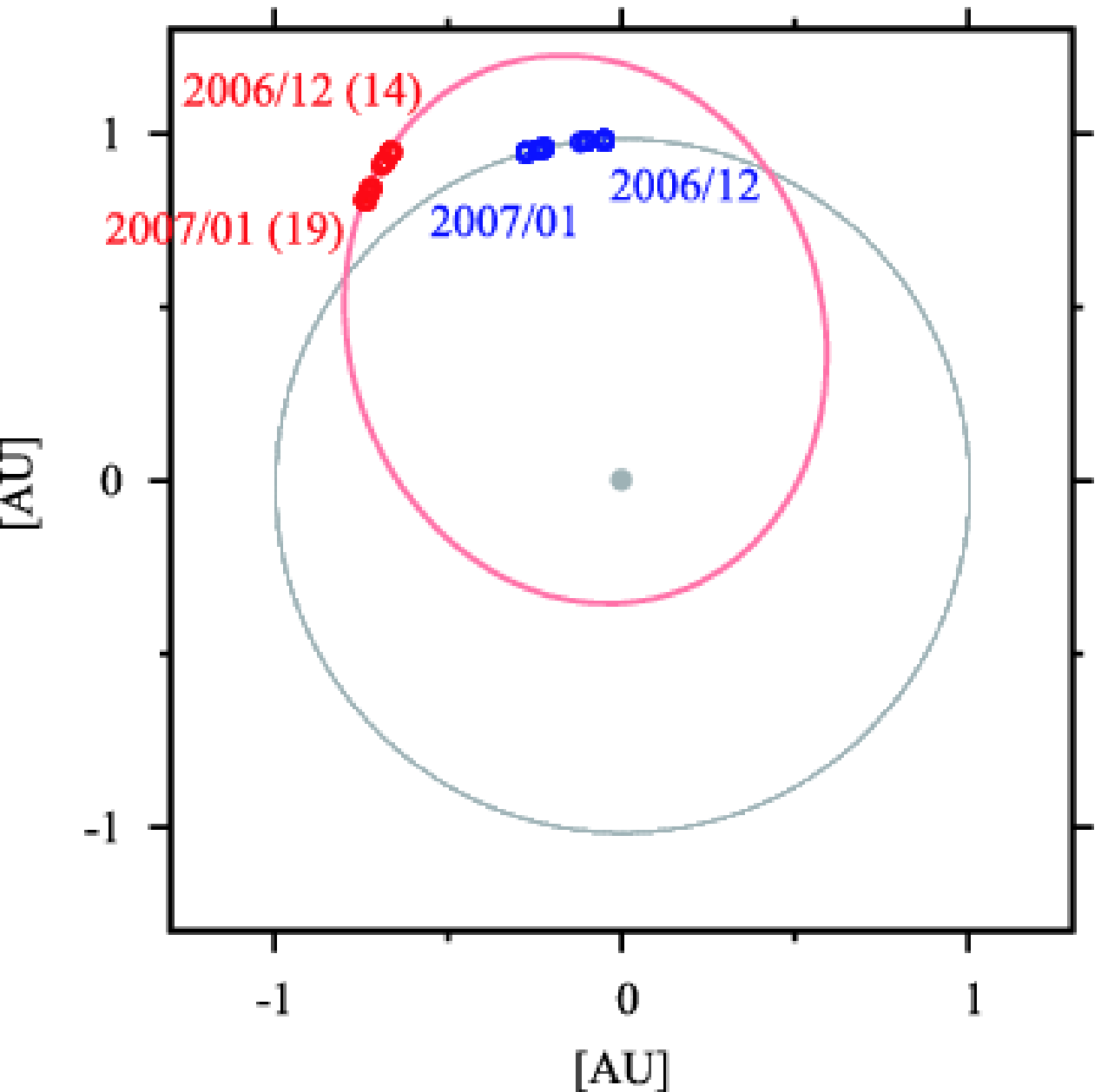}&%
\FigureFile(60mm,60mm){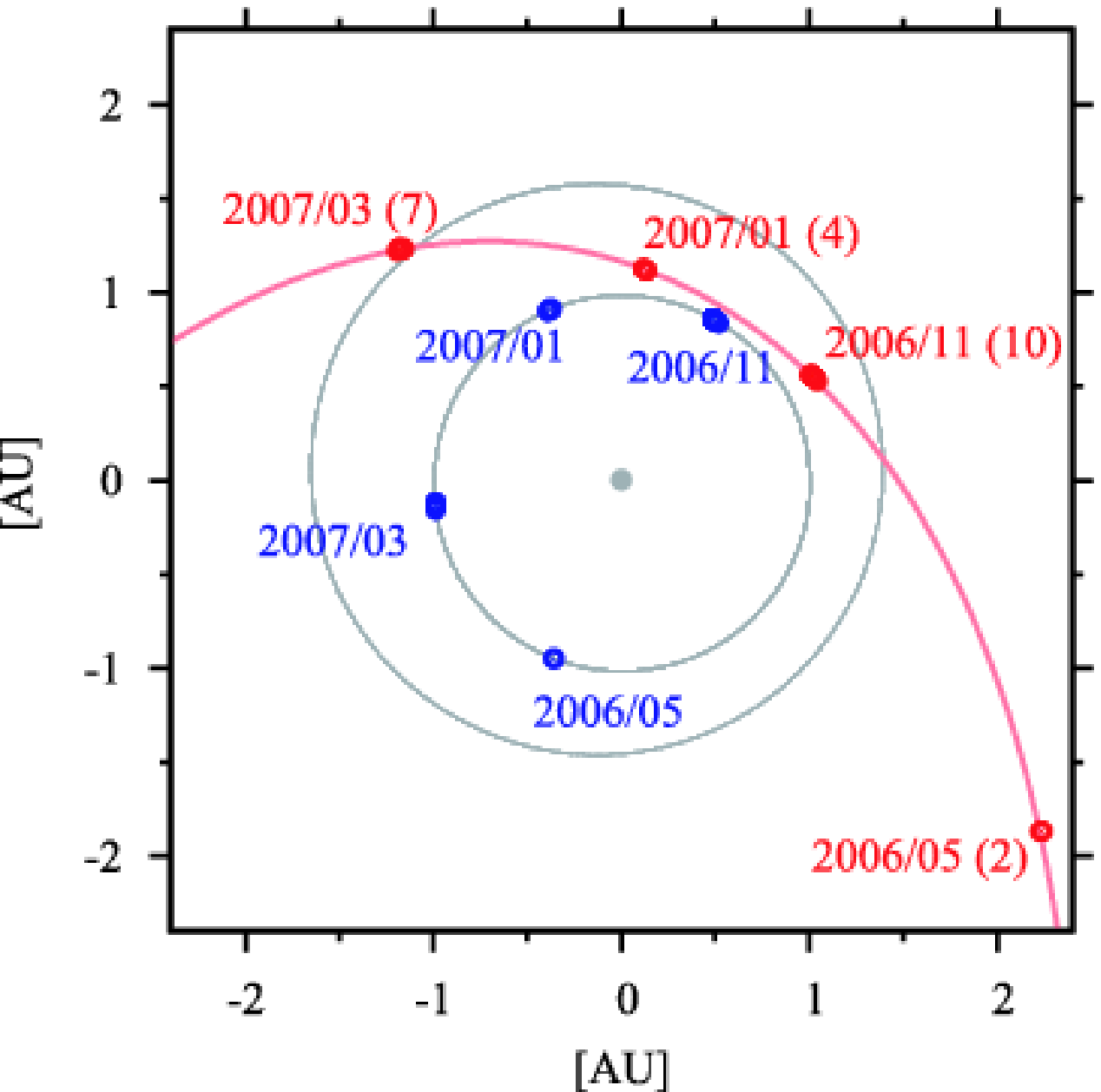}\\
\\
\hspace{20pt}85709 (1998 SG36): 22 detections & \hspace{20pt}366 Vincentina (1893 W): 16 detections\\[-0pt]
\FigureFile(60mm,60mm){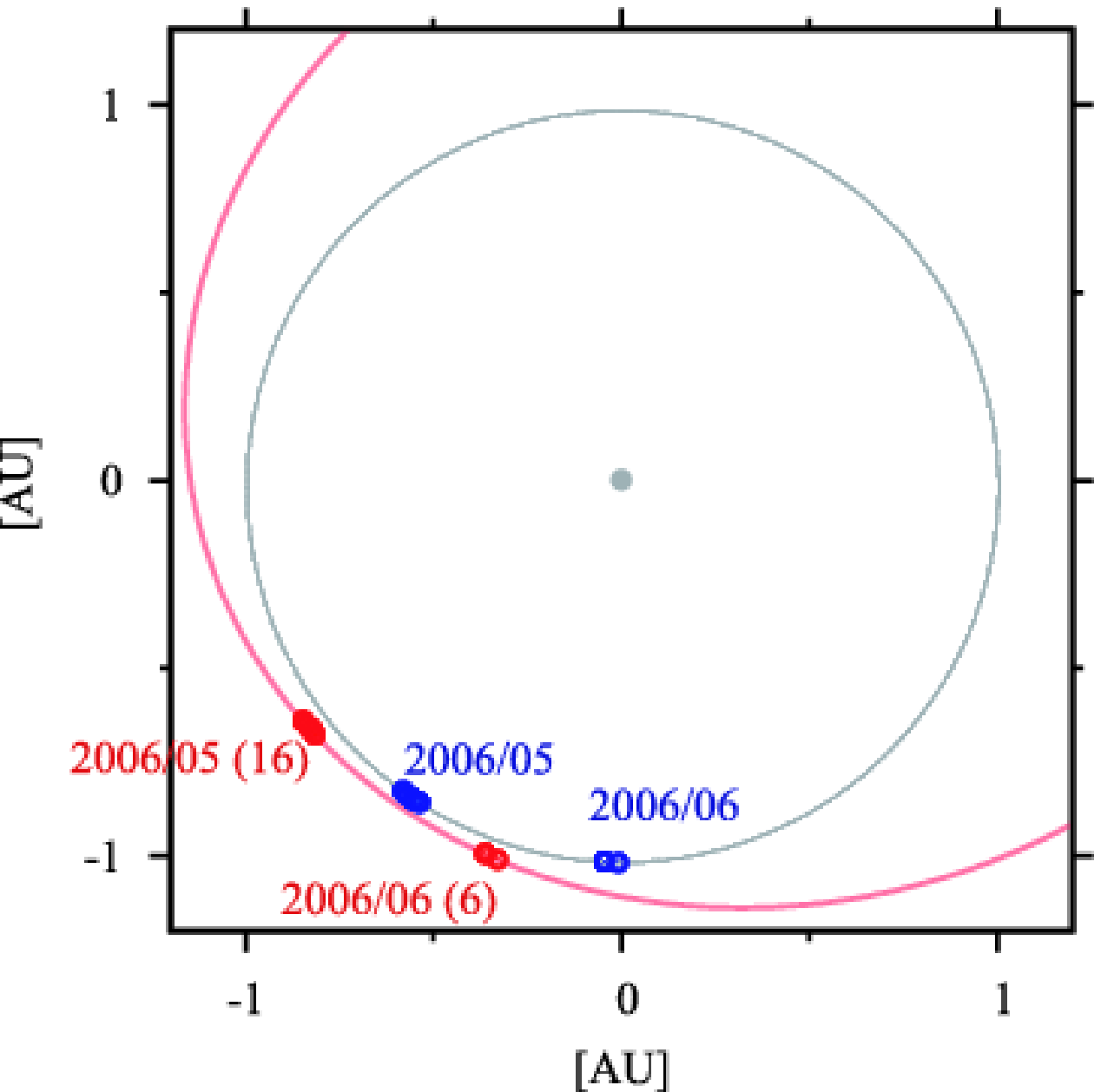}&%
\FigureFile(60mm,60mm){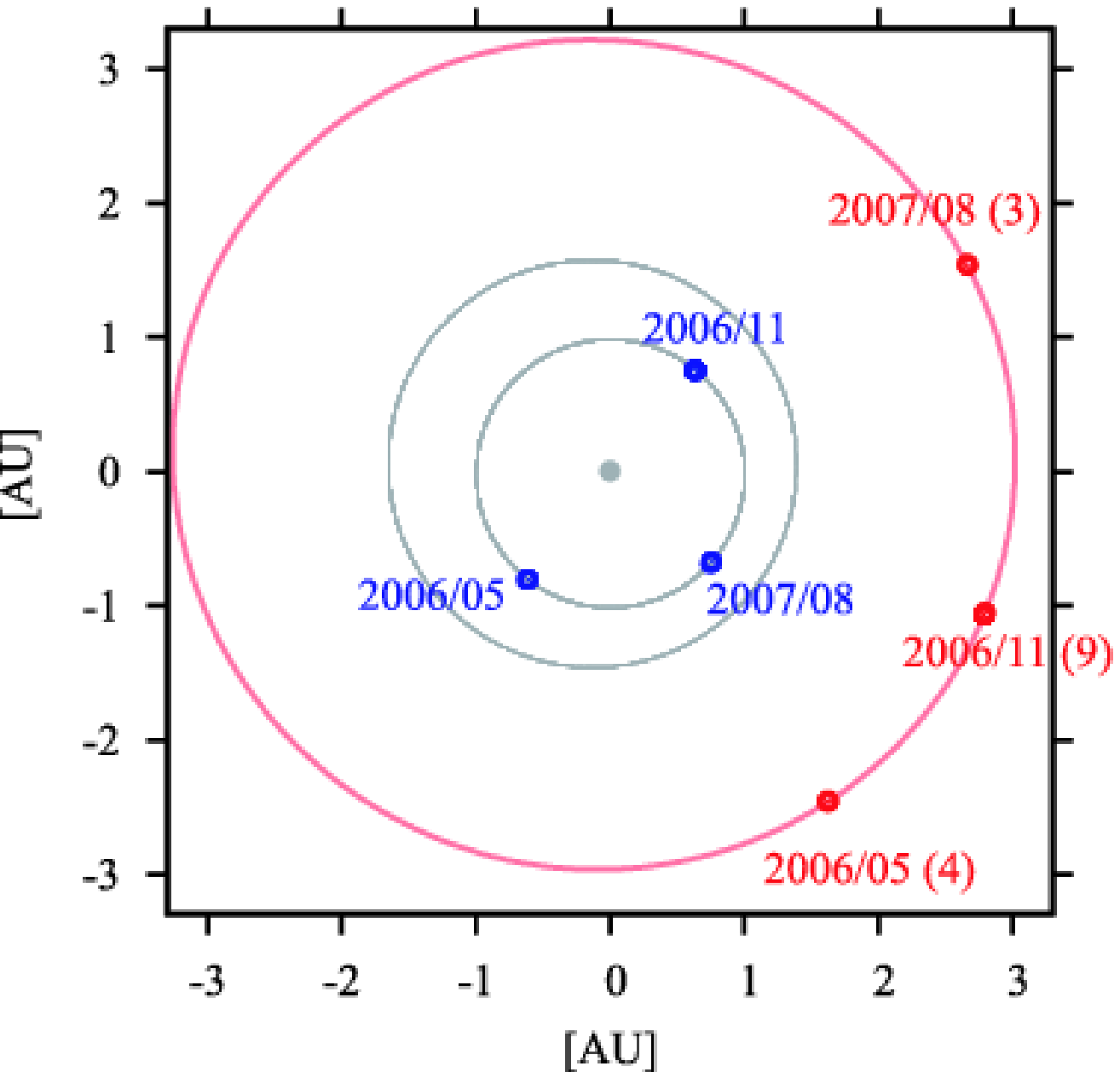}\\
\\
\hspace{20pt}7096 Napier (1992 VM): 15 detections & \hspace{20pt}7977 (1977 QQ5): 3 detections\\[-0pt]
\FigureFile(60mm,60mm){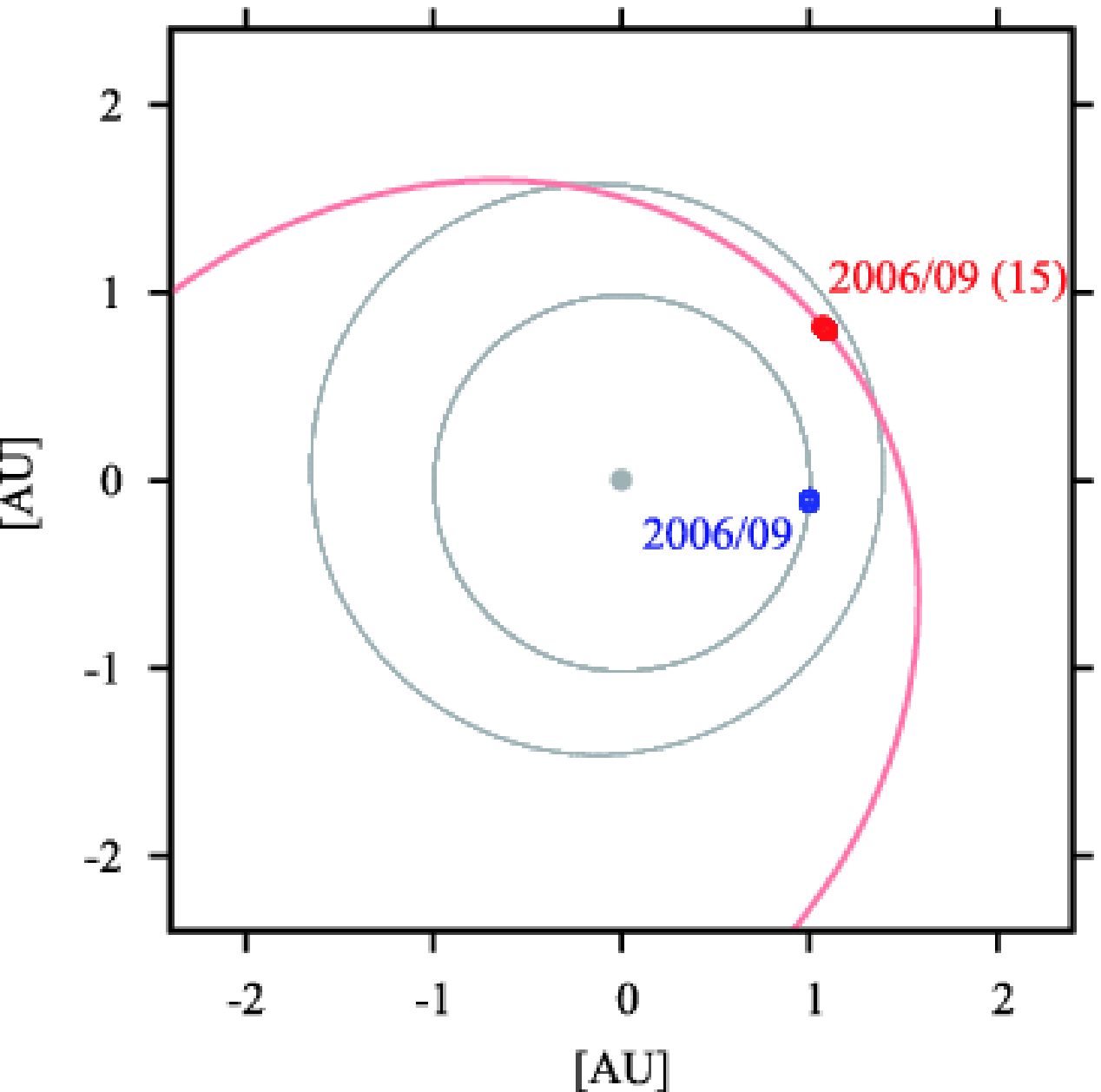}&%
\FigureFile(60mm,60mm){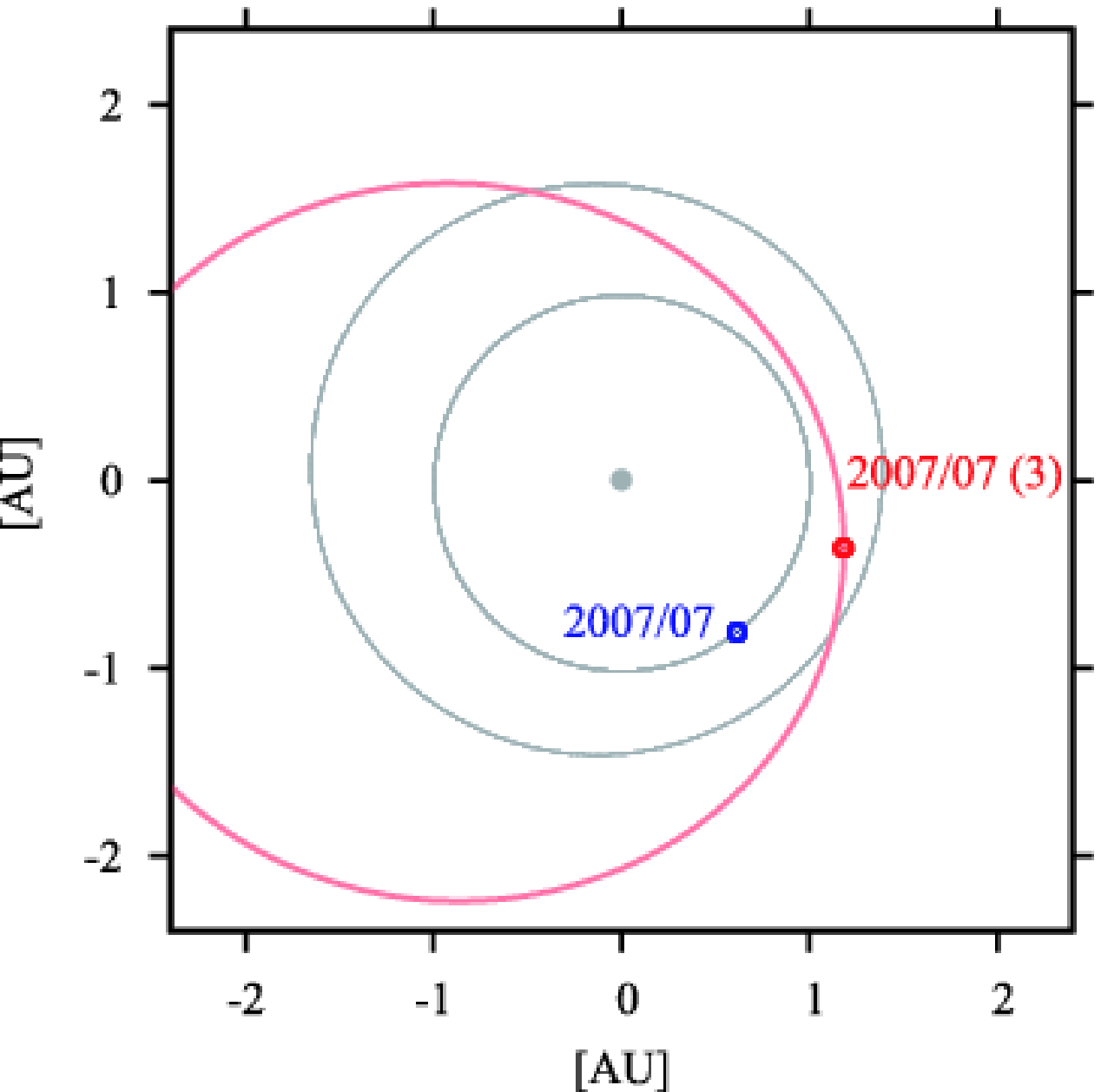}\\
\end{tabular}
}
\end{center}
\caption{
Orbits of the asteroids with the large number of detections projected on the ecliptic plane. 
7977 is an exceptional case in this figure (only 3 detections; see text). 
The red and blue open circles indicate the positions of the asteroids as of their detections, 
and those of the Earth, respectively.  The number of detections are given in the parentheses 
following the year/month of the observations.
The orientation is the same as Fig.\ref{fig:plot on ecliptic plane} but the scale is different.
}
\label{fig:plot on ecliptic plane for objects}
\end{figure*}

Fig.\ref{fig:histogram of NID} illustrates the number of detections 
for each asteroid with the AKARI All-Sky Survey. For comparison, we also plot the number 
of detections for the point sources in the IRC-PSC around the ecliptic plane, 
which includes galactic and extragalactic objects.  
AKARI 
basically observes a given portion of the sky at least twice in contiguous scans.  
Hence, a point source should have been observed four times at $S9W$ and $L18W$ in total.
Because the lifetime of the AKARI
cryogenic mission phase was 550 days, it observed a given portion of the sky in 
three different seasons.  Accordingly AKARI should have observed a point 
source on the ecliptic plane 12 times on average. The number could 
decrease because of the disturbance due to the SAA and the Moon
or increase by the offset survey described in 
Sect.\ref{The IRC All-Sky Survey by AKARI}. 
For the solar system objects, the situation is complicated 
due to their orbital motions. Considering the rate of change 
in the ecliptic longitude ($d\lambda/dt$), 
there are only five objects in AKARI catalog of 
$1.8^\prime/{\rm hr} < d\lambda/dt < 4.0^\prime/{\rm hr}$: 
137805 (2.96$^\prime$/hr), P/2006 HR30 (3.50$^\prime$/hr), 
85709 (2.95$^\prime$/hr), 7096 Napier (1.93$^\prime$/hr), 
and 7977 (2.66$^\prime$/hr), while the scan path of 
the All-Sky Survey shifts at most by $\sim$2.47$^\prime$/hr 
(= 360$^\circ$/yr) in the ecliptic longitude (i.e., 
in the cross-scan direction).  The orbits of these objects are 
illustrated in Fig.\ref{fig:plot on ecliptic plane for objects}. 
These objects except for 7977 have a large number of 
detections, e.g., more than 15 times, suggesting that they 
keep up with the scan direction: 33 times for 137805, 
23 times for P/2006 HR30, 22 times for 85709, and 
15 times for 7096 Napier. 
Although P/2006 HR30 is classified as a Halley-type comet (the Tisserand 
invariant value of $T_J = 1.785$; \citet{Hicks07}) and its cometary 
activity is reported (\cite{Lowry06}), we include this object as an 
asteroid in this paper. 
7977 has only 3 detections at $S9W$, due to the interference
with pointed observations as well as to the "negative" 
effect of the offset survey.  366 has 
$d\lambda/dt=0.49^\prime/{\rm hr}$, which is 
out of the range of the ``keep up'' speed mentioned above, but
it was observed 16 times. It has three
observation opportunities and at one of them (2006/11) 
the number of detections was increased by the "positive" 
effect of the offset survey. 

The present catalog contains only asteroids 
with the prograde motion with the Earth and no asteroids with the 
retrograde motion are included. The sources with multiple 
detection are more reliable in terms of the confirmation in general. 
The IRC-PSC only includes objects  that are 
detected at the same position at least twice.  The present catalog has
5,120 asteroids with $N_{\rm ID} \geq 1$, 
and 3,771 asteroids with $N_{\rm ID} \geq 2$, 
where $N_{\rm ID}$ is the number of the events with $S9W$ and $L18W$ in total.
It should be noted that the catalog includes asteroids with 
single detection ($N_{\rm ID} = 1$). 
The number of the detection is listed
in the catalog (Appendix \ref{catalog format}). 

\subsection{Size and Albedo distribution}
\label{Size and Albedo distribution}

\begin{figure}
\begin{center}
\FigureFile(80mm,80mm){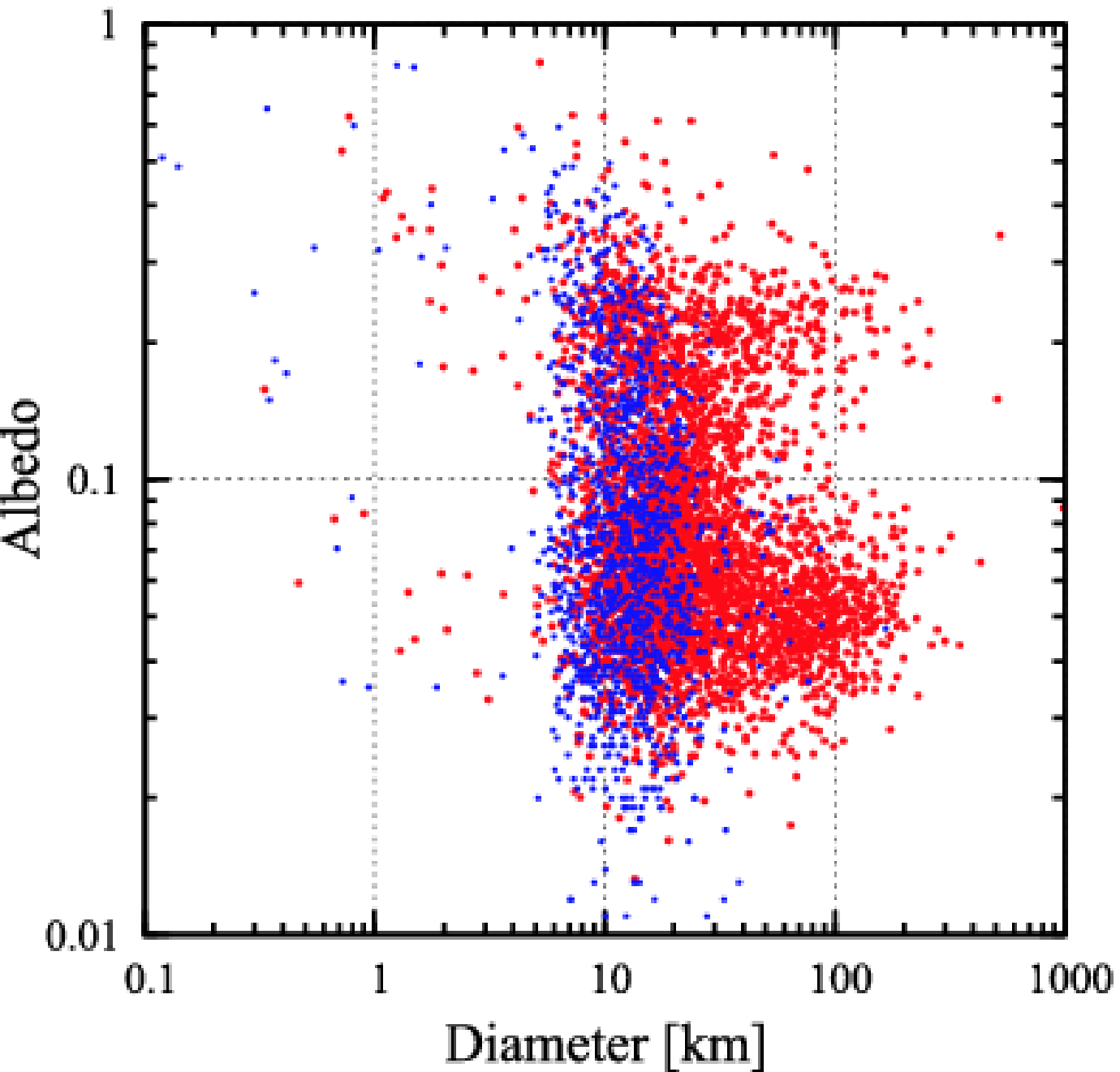}
\end{center}
\caption{Distribution of the size (diameter) and albedo of 
all the 5,120 identified asteroids.  
Red dots show those with more than two events and blue ones 
indicate those with single event detection.
}

\label{fig:diameter vs albedo}
\end{figure}

Fig.\ref{fig:diameter vs albedo} shows the distribution of albedos as a function of diameter
for the asteroids detected with the AKARI All-Sky Survey. An outstanding feature 
is the bimodal distribution in the albedo. It is also suggested 
that the albedo increases as the size decreases for small asteroids ($d<5$km),
although the number of the asteroids with 
the size of $d<5$km is not large. In the catalog, the smallest asteroid
is 2006 LD1, whose size is $d=0.12 \pm 0.01$km. The 
largest one is, naturally, 1 Ceres of $d=970 \pm 13$km. 

\begin{figure*}
\begin{center}
\begin{tabular}{cc}
\hspace{22pt}(a) & \hspace{22pt}(b)\\[-0pt]
\FigureFile(80mm,80mm){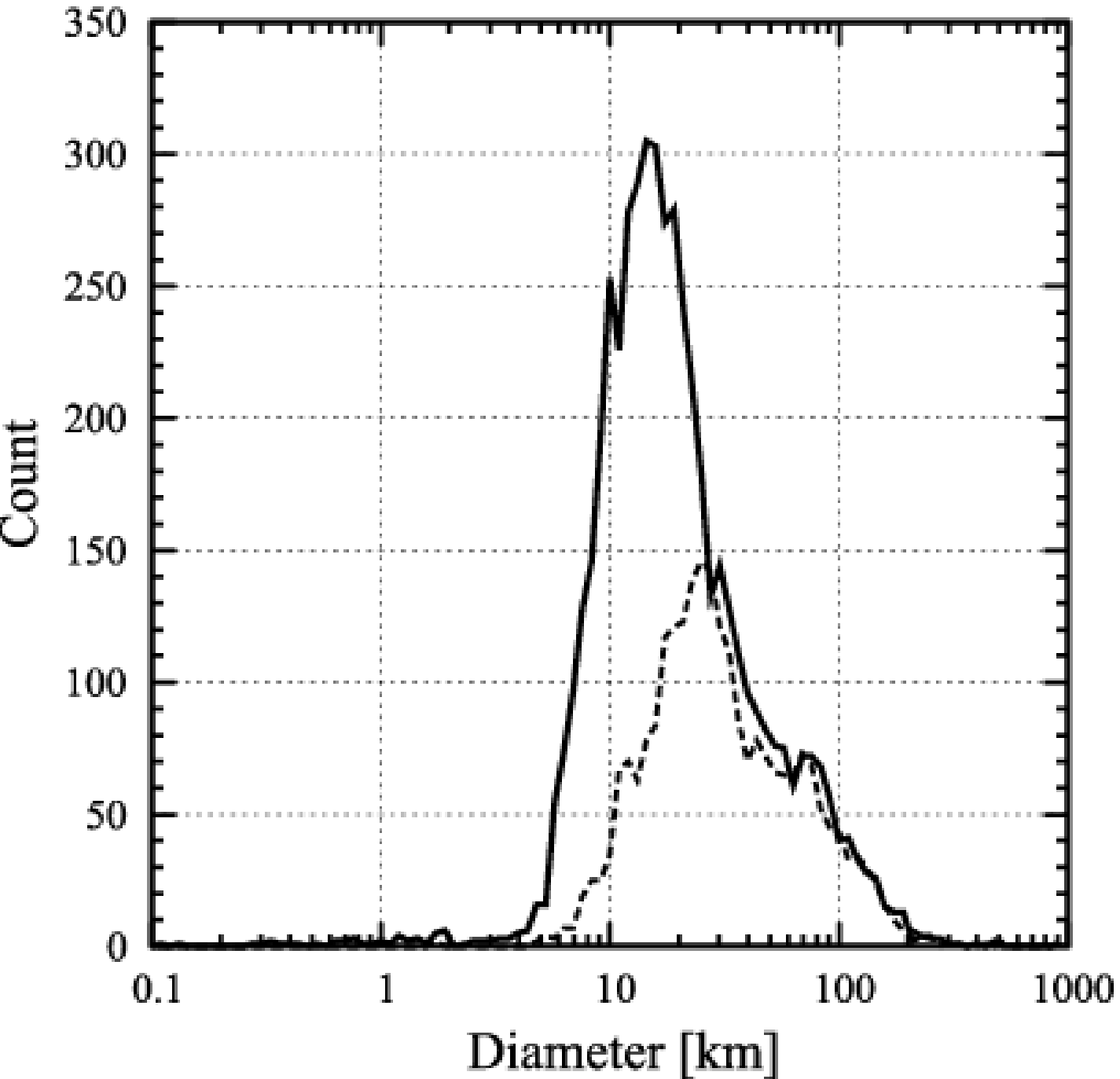} &%
\FigureFile(80mm,80mm){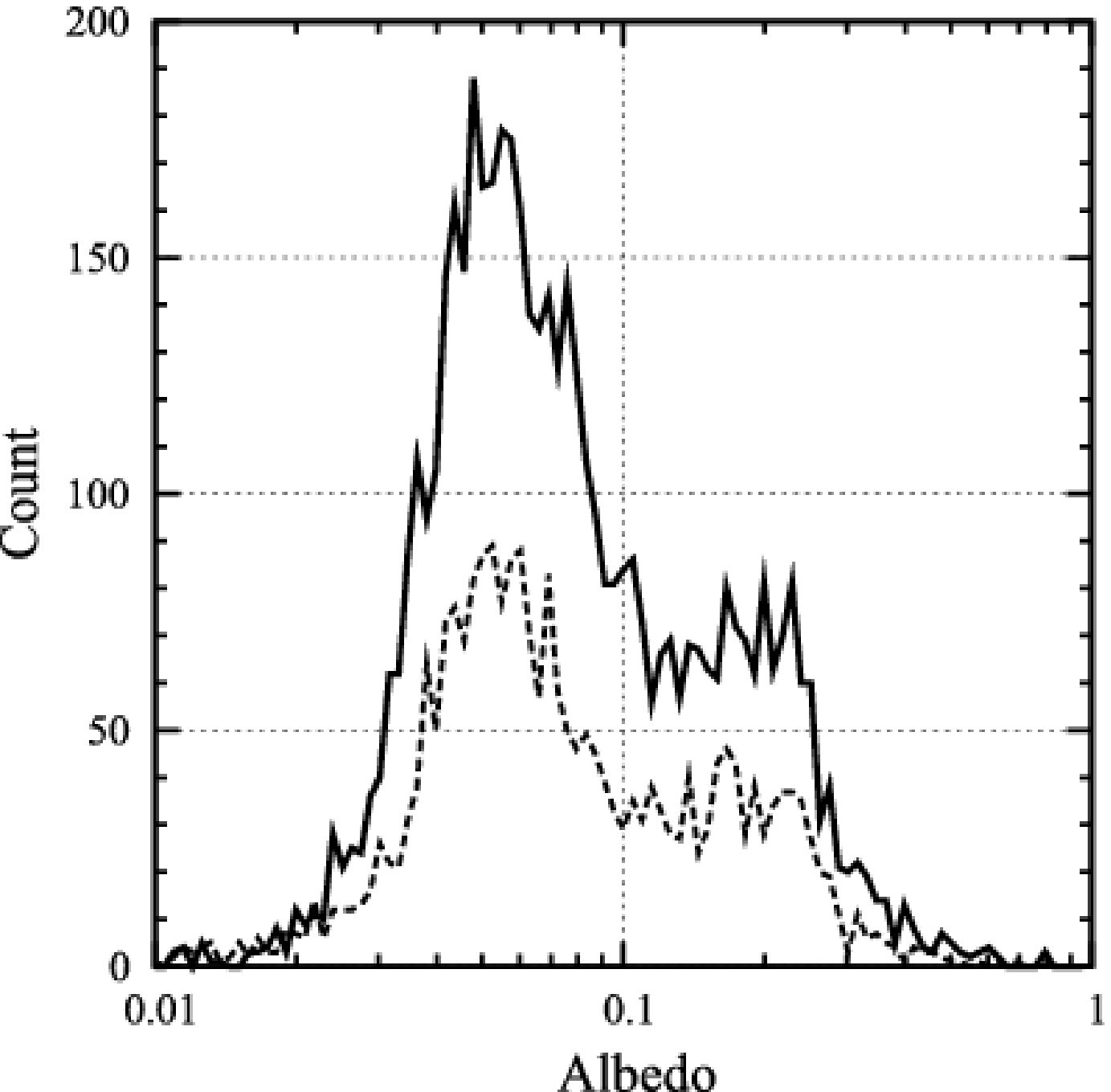}\\
\end{tabular}
\end{center}
\caption{Histograms of (a) the size (diameter) and (b) the albedo.
The solid, and dashed lines indicate the results from AKARI, 
and IRAS observations (\cite{SIMPS}), respectively.
The bin size is set as 100 segments for the range of 0.1km to 
1000km in the logarithmic scale for (a) and 100 segments for the range of 0.01 to 
1.0 in the logarithmic scale for (b). 
}
\label{fig:histogram of diameter and albedo}
\end{figure*}

Fig.\ref{fig:histogram of diameter and albedo} 
illustrates the histograms of the asteroids detected with the AKARI All-Sky Survey 
as a function 
of the size and the albedo.  For comparison, the 
results of IRAS observations are also plotted.  
The IRAS catalog consists of 2,228 objects with multiple detections 
and 242 objects with single detection (at 12\micron\, band). 
It clearly indicates that the AKARI All-Sky Survey
is more sensitive to small asteroids
than IRAS. Concerning about the size distribution of asteroids, the 
number is supposed to increase monotonically as the decrease of the size. 
Fig.\ref{fig:histogram of diameter and albedo}(a), however, shows
maxima around $d=15$km for AKARI and 30km 
for IRAS. The profiles of the histogram are similar to each other for those
larger than 30km, suggesting that IRAS and AKARI 
exhaustively detect asteroids of the size $d>30$km 
and $d>15$km, respectively, but that the completeness rapidly 
drops for asteroids
smaller than these values.  We discuss further on the size distribution
in the following section. 
Fig.\ref{fig:histogram of diameter and albedo}(b) clearly indicates that the albedo of
the asteroids has the well known bimodal distribution (\cite{Mo1977b}).  
The bimodal distribution can be attributed to two groups of taxonomic types of the asteroids. 
The primary peak around $p_{\rm v}=0.06$ is associated with C- and other low albedo types 
and the secondary peak around $p_{\rm v}=0.2$ with S- and other types with moderate albedo. 
Further discussion concerning about the taxonomic types will be discussed 
in a forthcoming paper (Usui et al. in preparation). 

\subsection{V-band magnitude of the identified asteroids}

\begin{figure*}
\begin{center}
\begin{tabular}{cc}
\hspace{20pt}(a) & \hspace{20pt}(b)\\[-0pt]
\FigureFile(80mm,80mm){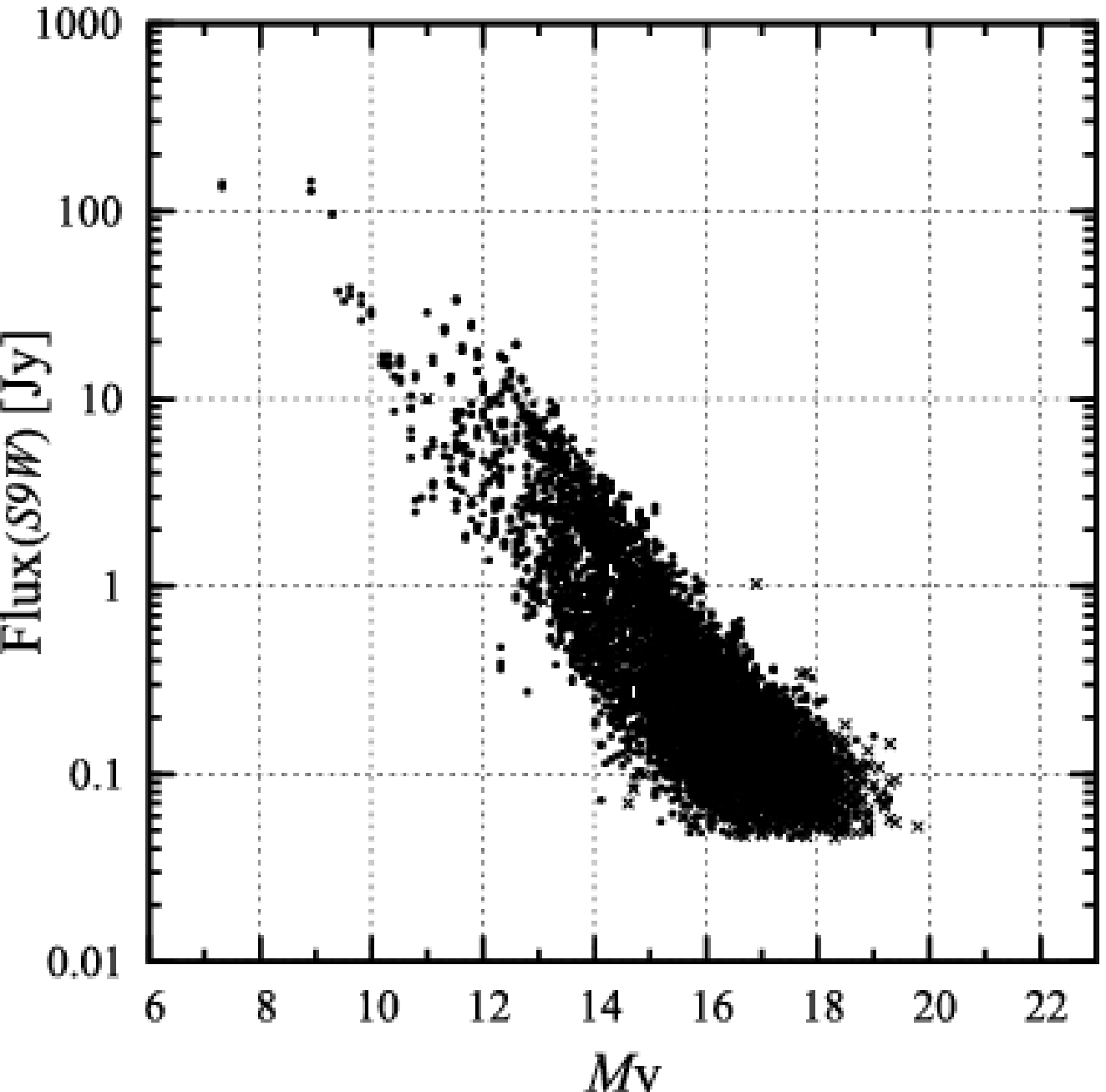} &%
\FigureFile(80mm,80mm){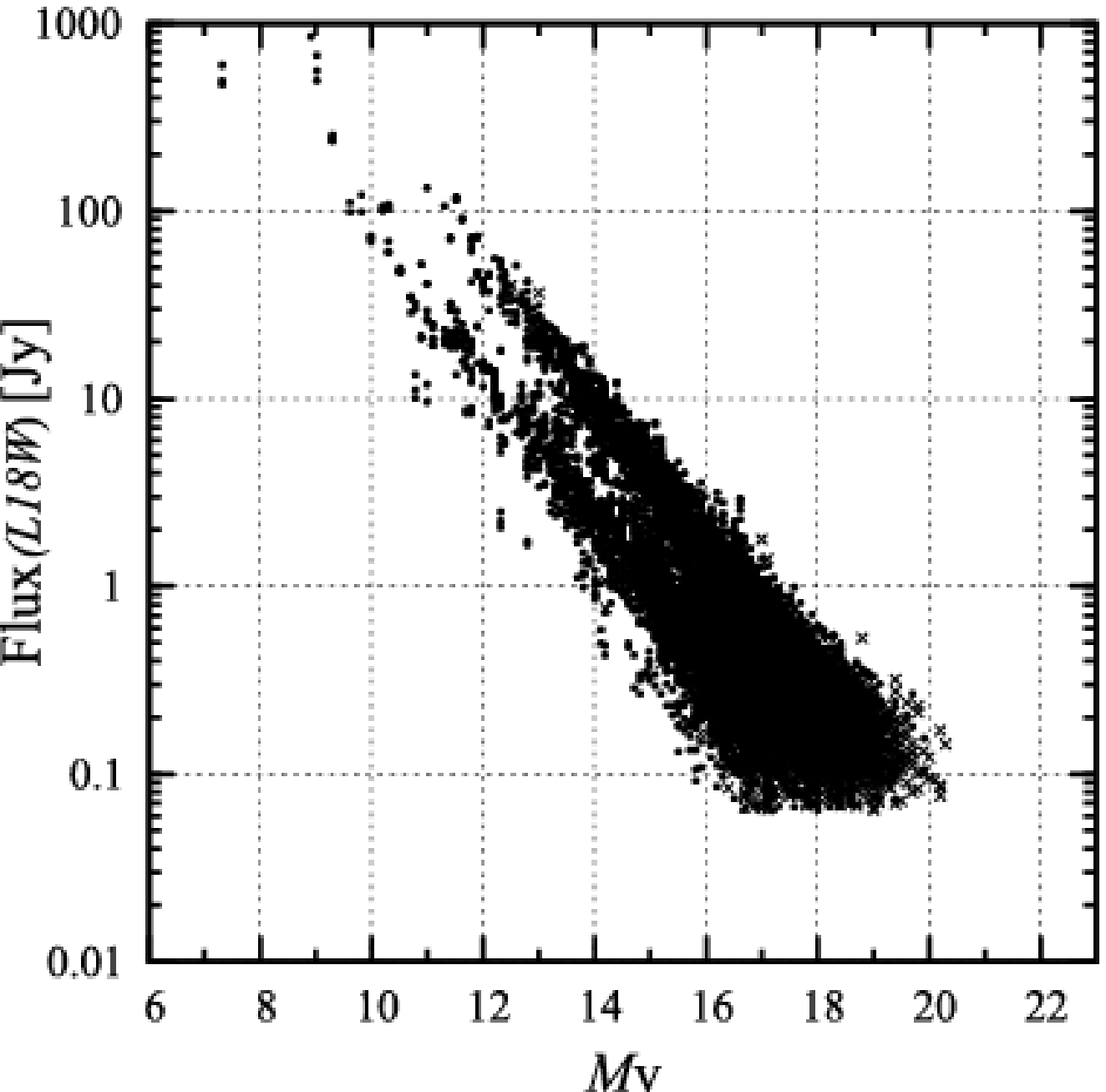} \\
\end{tabular}
\end{center}
\caption{Calculated $V$-magnitude ($M_V$) vs. color-corrected (monochromatic) flux of the
events identified as asteroids at $S9W$ and $L18W$.}
\label{fig:vmag vs flux}
\end{figure*}

Fig.\ref{fig:vmag vs flux} shows the calculated $V$-band magnitude ($M_V$) against 
the color-corrected monochromatic flux of the events 
identified as asteroids. 
3,771 asteroids have multiple 
events in the AKARI All-Sky Survey.  For example, 
4 Vesta is observed with the flux of 134--139 Jy at $S9W$ (2 times) 
and 474--604 Jy at $L18W$ (3 times) with $M_V = 7.3$,
1 Ceres is observed with the flux of 127--142 Jy at $S9W$ (3 times) and
497--853 Jy at $L18W$ (4 times) with $M_V = 8.9-9.0$, and
7 Iris is observed with the flux of 37--96 Jy at $S9W$ (3 times) and 
238--254 Jy at $L18W$ (4 times) with $M_V = 9.3-9.4$.
The bimodal characteristic is also seen in 
Fig.\ref{fig:vmag vs flux}. A sharp cutoff of the flux at below $\sim$ 0.1Jy is 
the result of the rejection of faint objects in the catalog 
processing (Sect.\ref{Statistical processing}).

We set a threshold for $M_V$
in the identification process (in step 7 in 
Sect.\ref{Asteroid identification}). 
The objects of the faintest $M_V$ in Fig.\ref{fig:vmag vs flux} are
67999 (2000 XC32) with $M_V=19.8$ at $S9W$ and 
102136 (1999 RO182) with $M_V=20.3$ at $L18W$.  It should be noted that
both objects are observed only once in the AKARI All-Sky Survey. This result confirms
that the threshold of $M_V=23$ in 
Sect.\ref{Asteroid identification} is reasonable to select real asteroids.


\subsection{Detection limit of the size of asteroids}
Fig.\ref{fig: diameter, albedo vs heldis} 
shows the estimated size of the asteroids as
a function of the heliocentric distance at the epoch of AKARI 
observation.  It is reasonable that smaller asteroids are
detected more in near-Earth orbits. 
No asteroids are detected inside of the Earth orbit because the viewing direction of the 
AKARI is fixed at the solar elongation of 90 $\pm$ 1 degree. The 
smallest asteroids detected around the Earth orbit, the outer main-belt 
(3.27AU), and Jupiter's orbit (5.2AU, Trojans) are 0.1km, 15km, and
40km, respectively.  

\begin{figure}
\begin{center}
\FigureFile(80mm,80mm){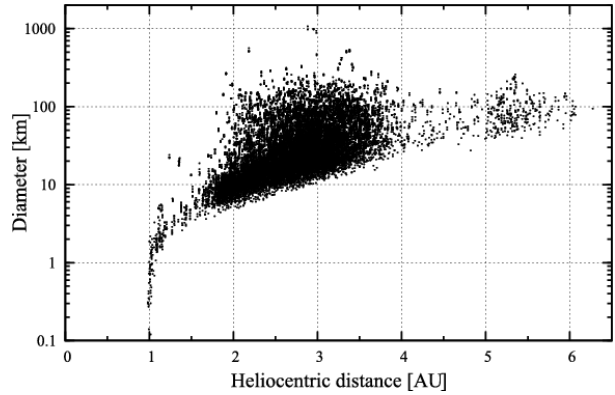}
\end{center}
\caption{Distribution of the estimated size (diameter)
vs. the heliocentric distance for the 
detected asteroids at the epoch of the observation with AKARI.}
\label{fig: diameter, albedo vs heldis}
\end{figure}


\subsection{Possibility of discovery of new asteroids}

In the asteroid catalog processing,
we did not take into account detection of new asteroids whose orbital
parameters are not known. Reliable detection of unknown moving objects 
require a high redundancy in the observation, which the AKARI All-Sky Survey did not
provide.  
Unfortunately the low visibility for observations
around the ecliptic plane 
makes it difficult to reliably detect new asteroids solely 
from the AKARI All-Sky Survey database.
However it is also very likely that the AKARI All-Sky Survey database contains
signals of undiscovered asteroids.  
In fact, we belatedly 
found that some asteroids had been detected with AKARI before their discovery.
For instance, 
2006 SA6, which was discovered on 2006 September 16 (\cite{Christensen2006}), had been detected 
on 2006 June 25 with AKARI, and 2007 FM3, which was discovered on 2007 March 19 (\cite{Kowalski2007}), had 
been observed on 2007 February 16 with AKARI 
(discovery of these two were done by Catalina Sky Survey). 
Thus whenever a new asteroid is discovered, we could check the detection in the AKARI All-Sky Survey database.  
\subsection{Comparison with previous work}

\subsubsection{Total number of detections}
\begin{figure}
\begin{center}
\FigureFile(80mm,80mm){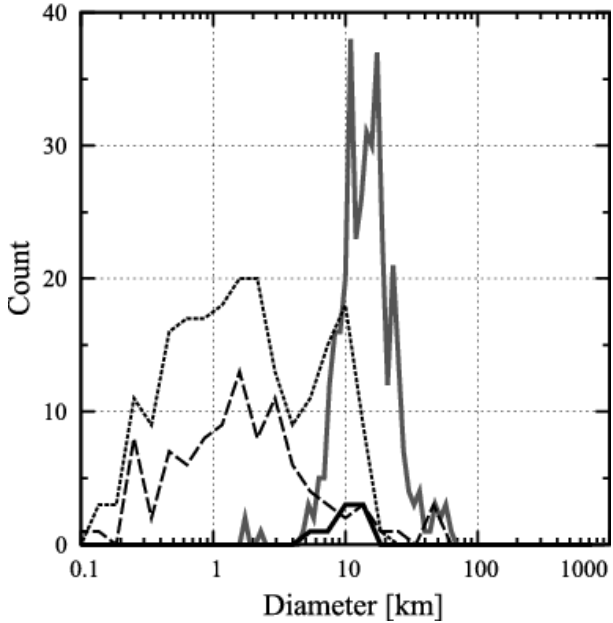}
\end{center}
\caption{Histogram of the asteroids with the previously determined size (diameter) 
without AKARI observations.  The gray solid, 
black solid, black dotted, and black dashed lines indicate the data 
with IRAS, MSX, SST, and other observatories, respectively.
The references are summarized in Appendix \ref{Reference list}. 
The bin size is set as 30 segments for the range of 0.1km to 
1000km in the logarithmic scale except for the data with IRAS, for which 
the bin size is set as 100 segments.}
\label{fig:AKARI failed asteroids}
\end{figure}

The total numbers of the detected asteroids with AKARI and previous work 
are summarized in Table \ref{table:number summary}.  The detected asteroids with AKARI 
are about twice as many as that of IRAS. A few hundreds of asteroids are not detected 
with AKARI, which have been observed previously. 
Fig.\ref{fig:AKARI failed asteroids} shows the size distribution of 
the asteroids undetected with AKARI.  Most observations of these asteroids were
made with 
the Spitzer Space Telescope (SST) and ground-based telescopes in programs to detect
small asteroids. 
Fig.\ref{fig:AKARI failed asteroids} indicates that AKARI 
All-Sky Survey did not detect hundreds of small asteroids of $d<15$km due the sensitivity
limit. 

\begin{table*}
\caption{Number of asteroids with derived radiometric 
size/albedo information. AKARI catalog compared to 
IRAS (\cite{SIMPS}), MSX (\cite{MIMPS}), 
SST (summarized in Appendix \ref{Reference list}:C1--C10), 
and other observations (in Appendix \ref{Reference list}:D1--D67).}
\label{table:number summary}
\begin{center}
\begin{tabular}{c|r|r|r|r|r}\hline
 & AKARI & IRAS  & MSX & SST & others \\\hline
asteroids with AKARI observations    & 5,120         & 2,103 & 160 &   7 & 288    \\
asteroids without AKARI observations &   ---         &   367 &   8 & 211 &  97    \\
total         & 5,120         & 2,470 & 168 & 218 & 385    \\\hline
\end{tabular}
\end{center}
\end{table*}

\subsubsection{Comparison with IRAS}
\label{Comparison with IRAS}

\begin{figure}
\begin{center}
\FigureFile(80mm,80mm){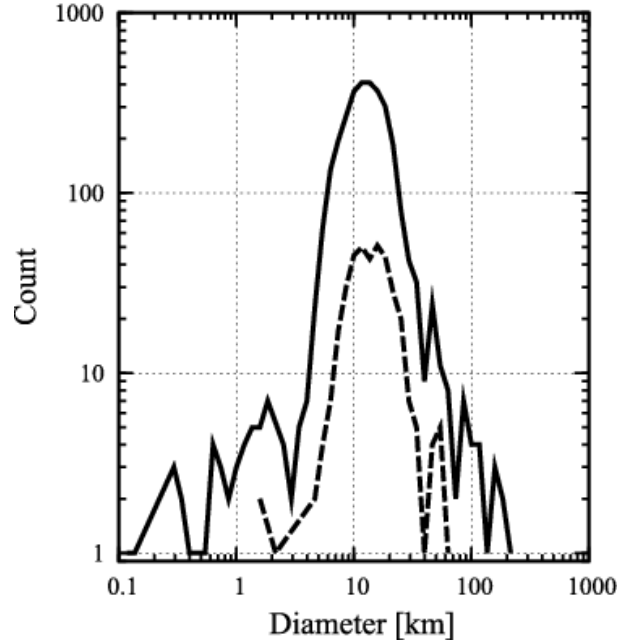}
\end{center}
\caption{
Histogram of the size (diameter) of the asteroids determined either by
AKARI or IRAS observations.  The solid and dashed lines indicate the numbers of the asteroids 
that are detected with AKARI but undetected with IRAS, and vice versa, 
respectively.  The bin size is set as 60 segments for the range of
0.1km to 1000km in the logarithmic scale.
}
\label{fig:comparison with IRAS}
\end{figure}

Fig.\ref{fig:comparison with IRAS} shows the histogram of the asteroids 
detected with AKARI without IRAS detection.  
A clear peak appears around the size of $d\sim 15$km, indicating 
that the AKARI All-Sky Survey extends the asteroid database down to $d\sim 15$km.
Table \ref{table:AKARI detections, without IRAS} shows large 
($d>100$km) asteroids detected with AKARI but undetected with IRAS.
Out of fifteen asteroids in this list, the size and 
albedo of three asteroids (375 Ursula (1893 AL), 190 Ismene, and 
275 Sapientia) have been determined by our measurements for the 
first time.  The size and albedo of the other twelve asteroids have been 
estimated with ground-based and space-borne telescopes previously. 
The AKARI asteroid catalog does not contain several very large ($d>40$km)
asteroids detected with IRAS
(Table \ref{table:IRAS detections, without AKARI}).  For these asteroids,
the size information is derived from IRAS observations.
All of these asteroids are distant objects and
belong to the Jovian Trojans except for 3 Hildas: 
11542 (1992 SU21), 4317 Garibaldi (1980 DA1), and 13035 (1989 UA6).
The semi-major axes of these objects are larger than 3.9 AU.  The closest 
heliocentric distances are about 5.1AU for the Trojans and 4.4AU for
the Hildas at the time of the AKARI All-Sky Survey observation.

The largest asteroid that AKARI failed to detect is 22180 (2000 YZ), whose 
size is $d=64$km according to IRAS observations.  
We examine the original scan data for these undetected large asteroids
and confirm that
two asteroids, 22180 (2000 YZ) and 4317 (1980 DA1), can be seen in 
raw images of the All-Sky Survey data at $L18W$ only once. 
They are, however,
rejected because they are detected near the edge of the detector.  
The other two asteroids, 14268 (2000 AK156) and 11542 (1992 SU21), 
are confused with stellar objects
since they are located at galactic latitudes less than 1 degree at the
epoch of the observation.  
For the other asteroids, no particular reasons for non-detection are found.
Some of them may lose the observation opportunities due to the offset survey operation 
mentioned in Sect.\ref{The IRC All-Sky Survey by AKARI}.
Deformed shapes, if any, may account for the non-detection with AKARI.

\begin{table*}
\caption{List of the asteroids that are detected with AKARI but undetected with
IRAS ($d > 100$km, 15 objects). The parameters $d$, $p_{\rm v}$, $a$, $e$, and $i$ 
indicate the size (diameter), the albedo, the semi-major axis, the eccentricity,
and the inclination of the asteroids, respectively.  The 
references are summarized in 
Appendix \ref{Reference list}.  
The cited data refer to the underlined reference in the list.
For those with the asterisks, namely, 375 Ursula, 
190 Ismene, and 275 Sapientia, the AKARI data provide the first determination of 
the size and albedo. 
}
\label{table:AKARI detections, without IRAS}
{\scriptsize
\begin{center}
\hspace*{-40pt}\begin{tabular}{rll|rrr|rr|rrl}\hline
&&&
\multicolumn{3}{c|}{orbital elements}&
\multicolumn{2}{c|}{AKARI}&
\multicolumn{3}{c}{previous work}\\\cline{4-11}
\multicolumn{3}{c|}{Asteroid}& $a$[AU]    & $e$      & $i$[deg] & \multicolumn{1}{c}{$d$[km]} & \multicolumn{1}{c|}{$p_{\rm v}$} & \multicolumn{1}{c}{$d$[km]} & \multicolumn{1}{c}{$p_{\rm v}$} & References\\\hline
 624  & Hektor    & 1907 XM  & 5.23749517 &0.02237543& 18.181769 & 230.99 $\pm$  3.94 & 0.034 $\pm$ 0.001 & 239.20 & 0.041 & \underline{D45}, D58\\
 19   & Fortuna   &          & 2.44236038 &0.15765176&  1.572523 & 199.66 $\pm$  3.02 & 0.063 $\pm$ 0.002 & 201.70 & 0.064 & D3, D5, D7, D16, \underline{D55}\\
 375  & Ursula    & 1893 AL  & 3.12268315 &0.10721155& 15.949598 & 193.63 $\pm$  2.52 & 0.049 $\pm$ 0.001 & \multicolumn{1}{c}{---}    & \multicolumn{1}{c}{---}   & (*)\\
 190  & Ismene    &          & 3.98157898 &0.16462886&  6.166222 & 179.89 $\pm$  3.64 & 0.051 $\pm$ 0.003 & \multicolumn{1}{c}{---}    & \multicolumn{1}{c}{---}   & (*)\\
 24   & Themis    &          & 3.12872103 &0.13118619&  0.759515 & 176.81 $\pm$  2.30 & 0.084 $\pm$ 0.003 & 176.20 & 0.084 & D52, \underline{D55}\\
 9    & Metis     &          & 2.38647903 &0.12228869&  5.574494 & 166.48 $\pm$  2.08 & 0.213 $\pm$ 0.007 & 154.67 & 0.228 & \underline{B1}, D3, D5, D42, D52, D55\\
 14   & Irene     &          & 2.58571736 &0.16721133&  9.105428 & 144.09 $\pm$  1.94 & 0.257 $\pm$ 0.009 & 155.00 & 0.170 & \underline{D3}, D5\\
 884  & Priamus   & 1917 CQ  & 5.16616811 &0.12330089&  8.925189 & 119.99 $\pm$  2.13 & 0.037 $\pm$ 0.001 & 138.00 & 0.034 & \underline{D45}\\
 129  & Antigone  &          & 2.86777878 &0.21205688& 12.218688 & 119.55 $\pm$  1.42 & 0.185 $\pm$ 0.005 & 115.00 & 0.187 & \underline{D7}\\
 275  & Sapientia &          & 2.77846168 &0.16053249&  4.768788 & 118.86 $\pm$  1.76 & 0.036 $\pm$ 0.001 & \multicolumn{1}{c}{---}    & \multicolumn{1}{c}{---}   & (*)\\
 3451 & Mentor    & 1984 HA1 & 5.10303310 &0.07129302& 24.695344 & 117.91 $\pm$  3.19 & 0.075 $\pm$ 0.005 & 122.20 & 0.052 & \underline{D45}\\
 127  & Johanna   &          & 2.75462967 &0.06479782&  8.241708 & 114.19 $\pm$  1.52 & 0.065 $\pm$ 0.002 & 123.33 & 0.056 & \underline{B1}\\
 27   & Euterpe   &          & 2.34596656 &0.17303724&  1.583754 & 109.79 $\pm$  1.54 & 0.234 $\pm$ 0.008 & 118.00 & 0.110 & \underline{D3}, D5\\
 481  & Emita     & 1902 HP  & 2.74051861 &0.15484764&  9.837640 & 103.53 $\pm$  1.90 & 0.061 $\pm$ 0.003 & 113.23 & 0.050 & \underline{B1}\\
 505  & Cava      & 1902 LL  & 2.68527271 &0.24493942&  9.839062 & 100.55 $\pm$  1.24 & 0.063 $\pm$ 0.002 & 115.80 & 0.040 & \underline{D55}\\\hline
\end{tabular}
\end{center}
}
\end{table*}

\begin{table*}
\caption{List of the asteroids which are detected with IRAS, not with AKARI ($d > 40$km, 11 objects). 
The columns are the same as in Table \ref{table:AKARI detections, without IRAS}.
}
\label{table:IRAS detections, without AKARI}
{\small
\begin{center}
\begin{tabular}{rll|r|r|r|r|r|l}\hline
&&&
\multicolumn{3}{c|}{orbital elements}&
\multicolumn{3}{c}{previous work}\\\cline{4-9}
\multicolumn{3}{c|}{Asteroid}               & $a$[AU] & $e$ & $i$[deg] & \multicolumn{1}{c|}{$d$[km]} & \multicolumn{1}{c|}{$p_{\rm v}$} & References\\\hline
  22180  &                     & 2000 YZ    &   5.19497082 &0.07172030 &29.276964 & 64.18 &  0.052 & \underline{A1}\\
  18137  &                     & 2000 OU30  &   5.13890227 &0.01629281 & 7.655306 & 60.71 &  0.013 & \underline{A1}\\
  5027   &  Androgeos          & 1988 BX1   &   5.30195674 &0.06662613 &31.450673 & 57.86 &  0.092 & \underline{A1}\\
  5025   &                     & 1986 TS6   &   5.20547347 &0.07670562 &11.022628 & 57.83 &  0.064 & \underline{A1}\\
  14268  &                     & 2000 AK156 &   5.26980857 &0.09197994 &14.950850 & 57.54 &  0.037 & \underline{A1}\\
  6545   &                     & 1986 TR6   &   5.12777404 &0.05220160 &11.998200 & 56.96 &  0.055 & \underline{A1}\\
  11542  &                     & 1992 SU21  &   3.95030034 &0.24086237 & 6.876531 & 49.72 &  0.022 & \underline{A1}\\
  4317   &  Garibaldi          & 1980 DA1   &   3.98754535 &0.16071342 & 9.823735 & 49.50 &  0.050 & \underline{A1}\\
  13362  &                     & 1998 UQ16  &   5.20935393 &0.02839927 & 9.334905 & 48.21 &  0.048 & \underline{A1}\\
  13035  &                     & 1989 UA6   &   3.97417007 &0.16620234 & 3.640840 & 47.40 &  0.018 & \underline{A1}\\
  11351  &                     & 1997 TS25  &   5.26120450 &0.06567781 &11.570297 & 42.16 &  0.063 & \underline{A1}\\\hline
\end{tabular}
\end{center}
}
\end{table*}

\begin{figure*}
\begin{center}
\begin{tabular}{cc}
\FigureFile(80mm,80mm){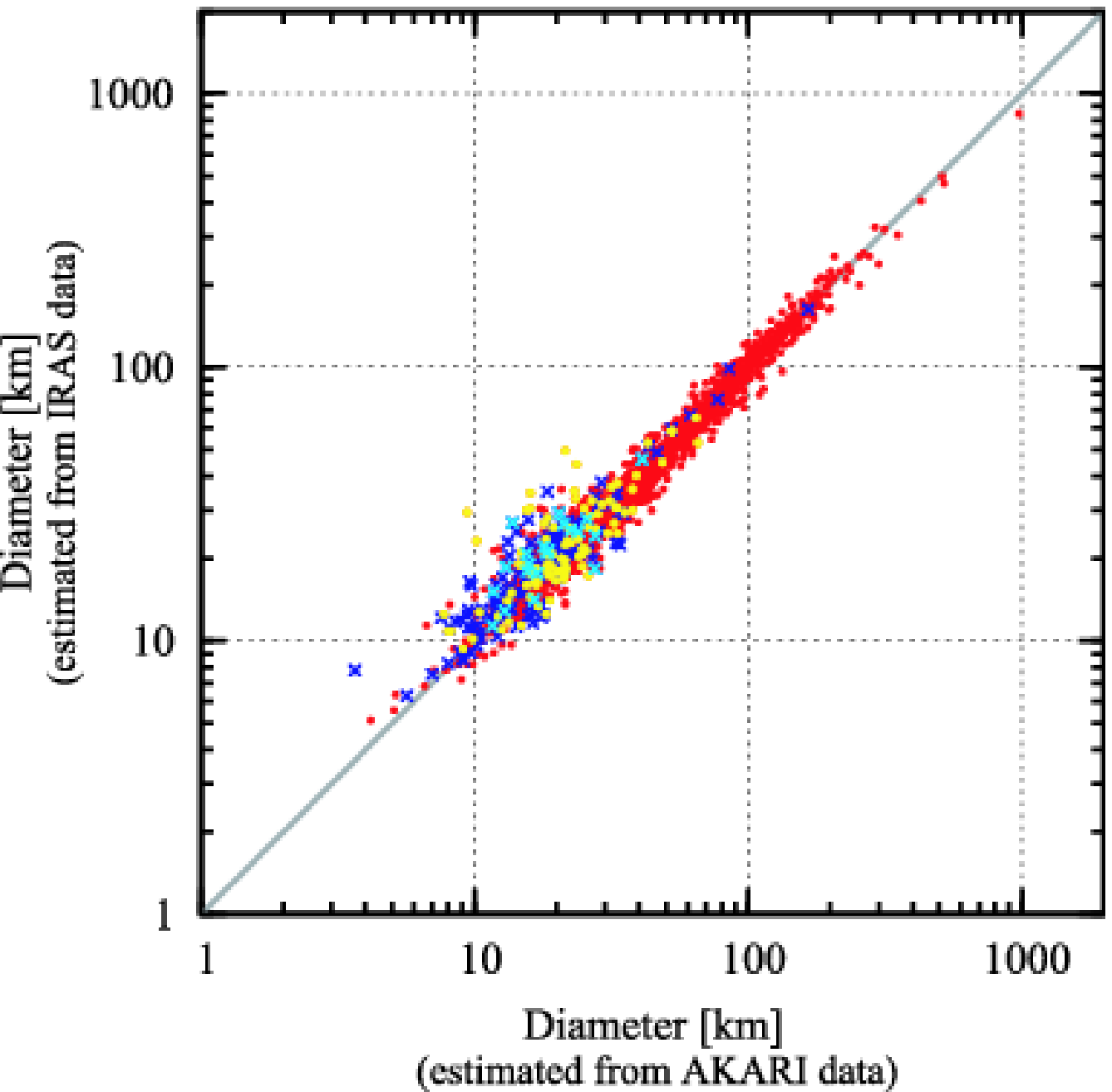} & %
\FigureFile(80mm,80mm){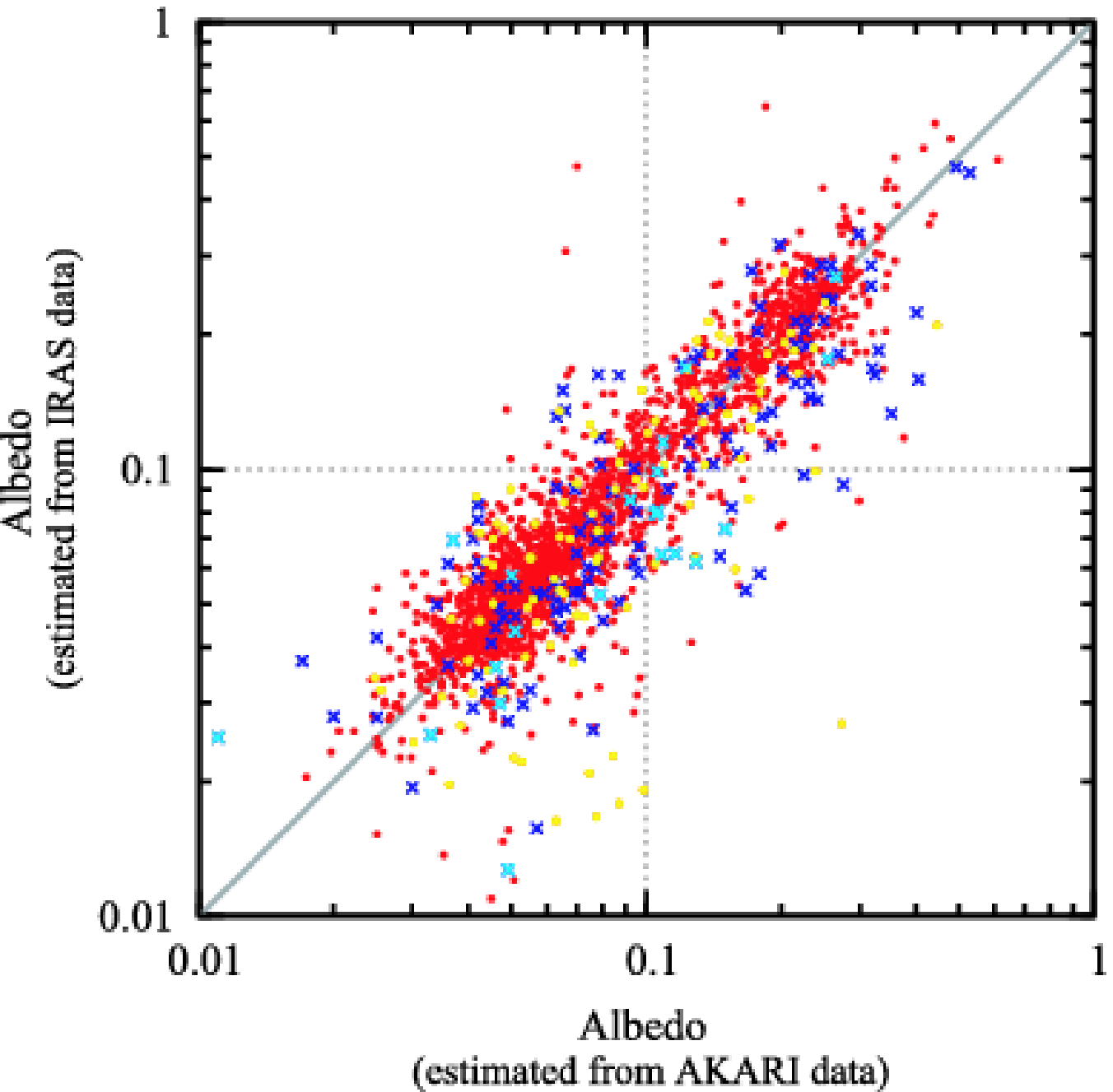}\\
&\\
\FigureFile(80mm,80mm){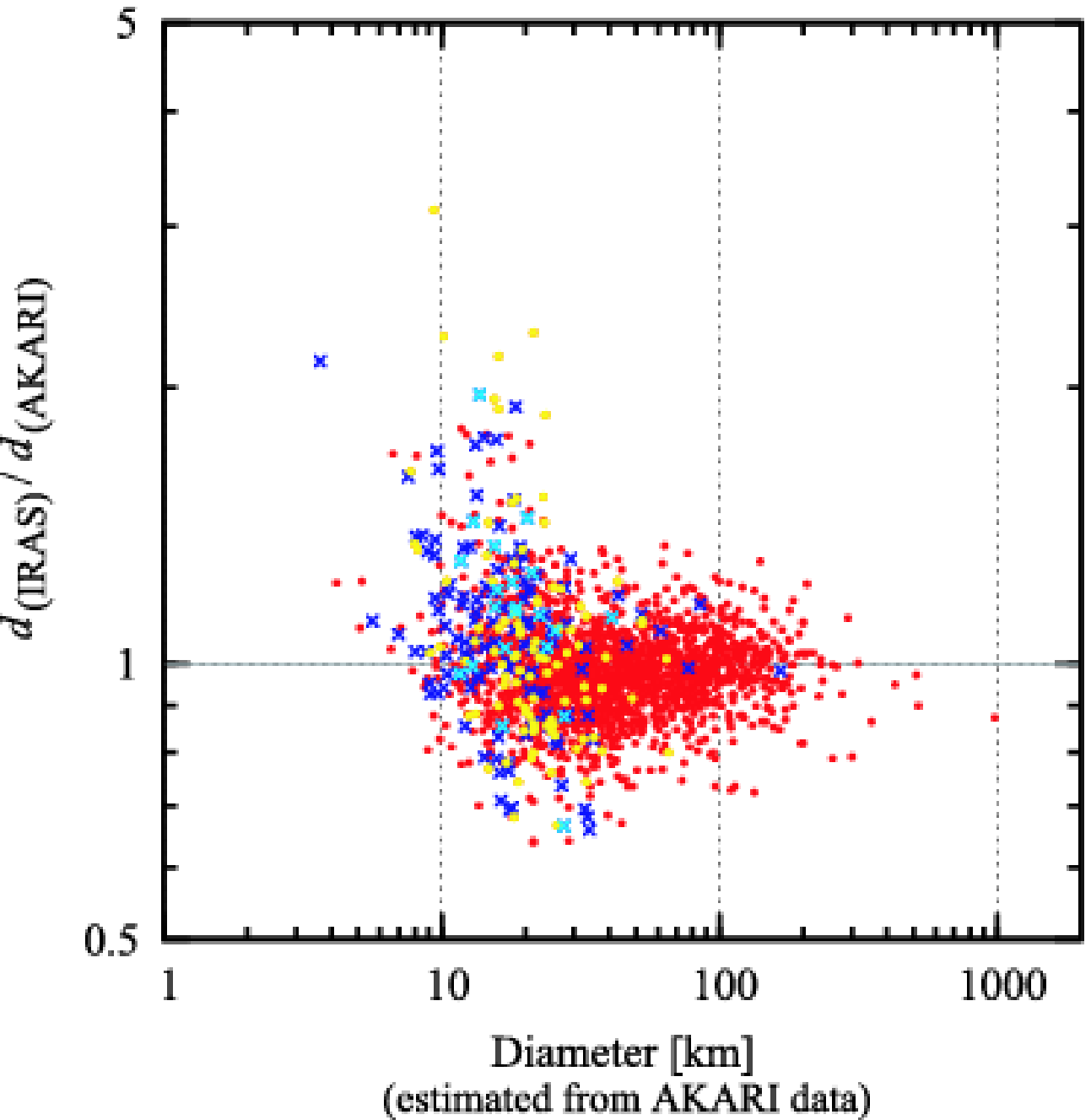} & %
\FigureFile(80mm,80mm){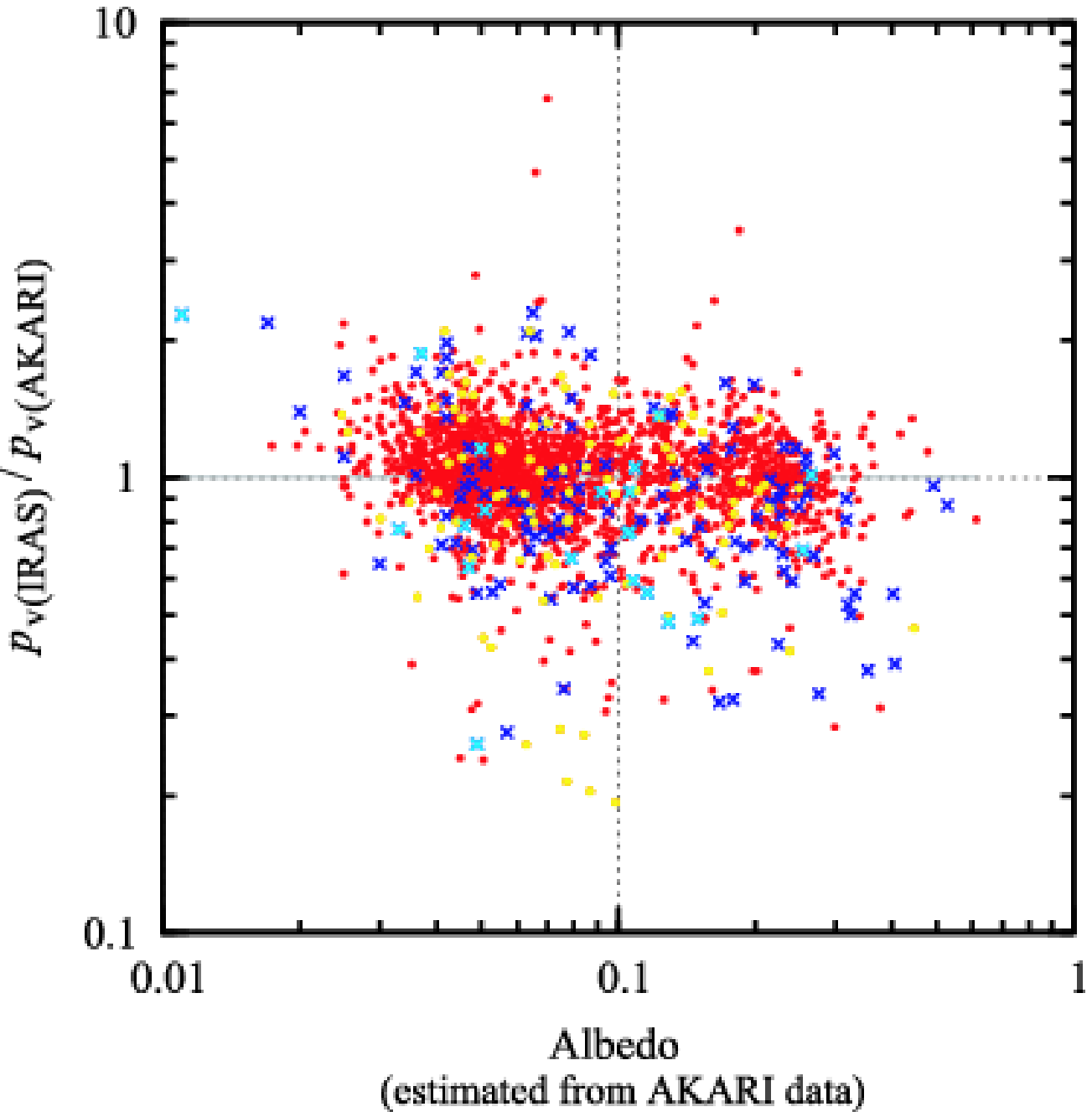}\\
\end{tabular}
\end{center}
\caption{Comparison between the estimates of AKARI and IRAS.
The number of the objects for each observation is shown in 
Table \ref{Number of objects between AKARI and IRAS}.  
The red dots, the yellow dots, the blue cross, and the light blue cross 
indicate the asteroids of (a), (b), (c), and (d) in 
Table \ref{Number of objects between AKARI and IRAS}, respectively. 
}
\label{fig:AKARI and IRAS}
\end{figure*}

Fig.\ref{fig:AKARI and IRAS} shows a comparison of 
the size and the albedo estimated from AKARI and IRAS observations for 
2,221 asteroids 
(Table \ref{Number of objects between AKARI and IRAS}). The two 
measurements show fairly good agreement.  The correlation coefficients 
are 0.9895 for the size and 0.8978 for the albedo for the asteroids 
observed twice or more (1,961 objects). 
However, there are several asteroids that show large discrepancies 
in their estimated size and albedo.  We list these asteroids in 
Table \ref{table:outliers with IRAS}.
The albedo of 1166 Sakuntala is estimated as 0.65 from IRAS and
$0.19 \pm 0.01$ from AKARI observations.  Because this
asteroid is classified as S-type, whose typical albedo is 0.216 
(or 0.158 in the Eight Color Asteroid Survey (ECAS) data 
(\cite{Zellner09}), see Table \ref{table:summary for 5 types}),
the estimate with AKARI is more likely to be correct.
Two asteroids 1384 Kniertje and 1444 Pannonia, have the albedo 
larger than 0.3 from IRAS but $\sim$0.07 from AKARI.  Since these two 
asteroids are of C-type (the mean albedo of 0.073, or 0.045 in ECAS), 
IRAS observations seem to overestimate the albedo.  The albedo of 5661 Hildebrand 
is estimated as 0.14 from IRAS and $0.049 \pm 0.003$ from AKARI observations.  Since 
this asteroid is a member of Hilda family composed of D-type asteroids
(\cite{dahlgren95}) which suggests the low albedo, the AKARI result seems to be more likely than IRAS.  

\begin{table}
\caption{Number of the asteroids for which the size and albedo are
estimated with AKARI and IRAS observations. 
$N_{\rm ID}$ indicates the number of the observations. They
are divided into four categories by $N_{\rm ID}=1$ or more
with AKARI and IRAS (a, b, c, and d) as shown in
Fig.\ref{fig:AKARI and IRAS}.}
\label{Number of objects between AKARI and IRAS}
\begin{center}
\begin{tabular}{c|c||r|r}\hline
\multicolumn{2}{c||}{}                  & \multicolumn{2}{c}{IRAS}\\\cline{3-4}
\multicolumn{2}{c||}{}                  & $N_{\rm ID} \geq 2$ & $N_{\rm ID} = 1$\\\hline\hline
AKARI & $N_{\rm ID} \geq 2$ & 1,961 (a)                      &   97 (b)\\\cline{2-4}
      & $N_{\rm ID} =    1$ &   142 (c)                      &   21 (d)\\\hline
\end{tabular}
\end{center}
\end{table}

\begin{table*}
\caption{Asteroids that show large discrepancy in the size and albedo estimated from IRAS and AKARI observations}
\label{table:outliers with IRAS}
\begin{center}
\begin{tabular}{rll|rrr|rrr|cc}\hline
\multicolumn{3}{c|}{}         & \multicolumn{3}{c|}{IRAS data} & \multicolumn{3}{c|}{AKARI data} & taxonomic &\\[-0pt]
\multicolumn{3}{c|}{Asteroid} & \multicolumn{1}{c}{$d$[km]} & \multicolumn{1}{c}{$ p_{\rm v}$} & $N_{\rm ID}$\footnotemark[\dag] & \multicolumn{1}{c}{$d$[km]} & \multicolumn{1}{c}{$ p_{\rm v}$} & $N_{\rm ID}$\footnotemark[\ddag] &type&family\\\hline
\multicolumn{3}{l|}{(Asteroids with discrepant size)} &&&&&&\\
 1293 & Sonja  & 1933 SO  &  7.80 & 0.460 & 3 &  3.65 $\pm$ 0.45 & 0.529 $\pm$ 0.133 & 1 & S  & ---\\
 5356 &        & 1991 FF1 & 29.37 & 0.027 & 1 &  9.39 $\pm$ 0.70 & 0.273 $\pm$ 0.044 & 2 & --- & ---\\
 7875 &        & 1991 ES1 & 34.58 & 0.018 & 1 & 15.95 $\pm$ 0.45 & 0.087 $\pm$ 0.005 & 5 & --- & ---\\
14409 &        & 1991 RM1 & 49.31 & 0.017 & 1 & 21.45 $\pm$ 0.88 & 0.077 $\pm$ 0.007 & 3 & X(P)  & ---\\
16447 & Vauban & 1989 RX  & 23.10 & 0.019 & 1 & 10.17 $\pm$ 0.70 & 0.098 $\pm$ 0.014 & 2 & --- & ---\\
&&&&&&&&\\
\multicolumn{3}{l|}{(Asteroids with discrepant albedo)} &&&&&&\\
 1166 & Sakuntala  & 1930 MA  & 28.74 & 0.646 & 5 & 26.32 $\pm$ 0.39 & 0.185 $\pm$ 0.006 & 8 & S  & ---\\
 1384 & Kniertje   & 1934 RX  & 27.51 & 0.308 & 8 & 26.14 $\pm$ 0.56 & 0.066 $\pm$ 0.003 & 7 & C  & Adeona\\
 1444 & Pannonia   & 1938 AE  & 29.20 & 0.475 & 2 & 30.48 $\pm$ 0.53 & 0.070 $\pm$ 0.003 & 7 & C(B)  & ---\\
 5661 & Hildebrand & 1977 PO1 & 34.37 & 0.136 & 2 & 42.29 $\pm$ 1.26 & 0.049 $\pm$ 0.003 & 5 & --- & Hilda\\\hline
\multicolumn{11}{l}{\hbox{\parbox{160mm}{\footnotesize
      \par\noindent
      \footnotemark[\dag] The number of the observations used in the estimate of the albedo. 
      \par\noindent
      \footnotemark[\ddag] The number of the detections with $S9W$ and $L18W$.
}\hss}}

\end{tabular}
\end{center}
\end{table*}

\begin{table*}
\caption{Summary for 5 taxonomic classes of the asteroids detected with
AKARI. 
}
\label{table:summary for 5 types}
\begin{center}
\begin{tabular}{l|rrr||rrr}\hline
\multicolumn{1}{c}{\footnotesize taxonomic} & %
\multicolumn{3}{|c||}{AKARI} &
\multicolumn{3}{|c}{ECAS}\\
\multicolumn{1}{c}{\footnotesize type }     & %
\multicolumn{1}{|c}{number} &%
\multicolumn{1}{c}{$\overline{p_{\rm v}}$}&%
\multicolumn{1}{c||}{$\sigma\left(\overline{p_{\rm v}}\right)$}&%
\multicolumn{1}{c}{number}&%
\multicolumn{1}{c}{$\overline{p_{\rm v}}$}&%
\multicolumn{1}{c}{$\sigma\left(\overline{p_{\rm v}}\right)$}\\\hline
\multicolumn{1}{c|}{C}            &     616  & 0.073  & 0.043 & 62 & 0.045 & 0.010 \\
\multicolumn{1}{c|}{S}            &     614  & 0.216  & 0.086 & 78 & 0.158 & 0.038 \\
\multicolumn{1}{c|}{X}            &     418  & 0.106  & 0.101 & 52 & 0.126 & 0.027 \\
\multicolumn{1}{c|}{D}            &     165  & 0.075  & 0.051 & 20 & 0.031 & 0.005 \\
\multicolumn{1}{c|}{V}            &       5  & 0.296  & 0.113 &  1 & 0.249 & \multicolumn{1}{c}{---} \\\hline
total        &    1818  &        &       &213 &       &       \\\hline
\multicolumn{7}{l}{\hbox{\parbox{100mm}{\footnotesize
      \par\noindent
Note: Current version of ECAS (Eight-Color Asteroid Survey; \cite{Zellner09})
contains not only the database of the reflectance spectrophotometric survey but also related dataset
including the geometric albedo which we refer in this Table. \\
The mean albedo is taken as the average value of the taxonomic classes that 
belong to the same taxonomic type, 
i.e., C-type: B, C, F, and G; S-type: A, Q, R, and S; X-type: X, M, and P; 
D-type: D and T; and V-type: V. \\
Determination of the taxonomic classes for AKARI samples is based on
the references summarized in Appendix \ref{Reference list for taxonomic types}:E1--E42.
}\hss}}
\end{tabular}
\end{center}
\end{table*}

The discrepancy in the size estimate needs more detailed investigation.
For 1 Ceres, the largest asteroid in the main belt 
or one of the dwarf planets, IRAS estimate the size as 850km and AKARI 
as $970 \pm 13$km. 
Hubble Space Telescope observations (\cite{thomas05}) derive it as 974.6 $\times$ 909.4km,
supporting the AKARI estimate.  For the other 5 asteroids listed 
in upper rows of Table \ref{table:outliers with IRAS} only have sizes 
determined with IRAS and AKARI and thus it is difficult to conclude which would be
more accurate. Further observations 
and measurements are needed to understand the discrepancy in size 
between IRAS and AKARI.


\section{Concluding remarks}

We have created an unbiased, homogeneous asteroid catalog, which 
contains a total of 5,120 objects. This is the second generation asteroid survey 
after IRAS.  The catalog revises the properties of several asteroids. 
The catalog will be publicly available via the Internet. 
This catalog will be significant for the various fields of the
solar system science and contribute to future Rendezvous and/or sample 
return missions of small objects.
\bigskip

This study is based on observations with AKARI, a JAXA project with
the participation of ESA. We would like to thank all the members of
AKARI project for their devoted efforts to achieve our observations. 
We are grateful to I. Yamamura for kindly providing us with the
computer system including the web server for the public release of 
our catalog. 
Thanks are due to the referee, Simon Green, for careful reading and 
providing constructive suggestions, which have greatly helped to 
improve this paper. 
FU would like to thank C. P. Pearson for his helpful comments. 
SH is supported by the Space Plasma Laboratory, ISAS/JAXA.
This work is supported in part by 
Grant-in-Aid for Scientific Research on Priority Areas No.19047003 to DK, 
Grant-in-Aid for Young Scientists (B) No.21740153 and Scientific Research on Innovative Areas No. 21111005 to TO, 
Grant-in-Aid for Scientific Research (C) No.19540251 to HK, and 
Grant-in-Aid for JSPS Fellows No.10J02063 to ST, 
from the Ministry of Education, Culture, Sports, Science 
and Technology of Japan.


\newpage

\appendix

\section{Format of the AKARI/IRC Mid-infrared Asteroid Survey}
\label{catalog format}
The Asteroid Catalog Using Akari (AcuA) will be publicly available on 
the web server at the Institute of Space and Astronautical Science (ISAS), 
Japan Aerospace Exploration Agency (JAXA)\footnote{http://www.ir.isas.jaxa.jp/AKARI/Observation/ (not yet prepared at the time of submission to astro-ph)}.
It contains 5,120 asteroids detected in the mid-infrared with
the size, albedo, and their associated uncertainties. 
It is in a standard ASCII file with a fixed-length record. 
Each line corresponds to each object with 10 columns.
A summary of the format is given in Table \ref{table:catalog contents}. 
NUMBER, NAME, and PROV\_DES are the asteroid number, the name, and the
provisional designation, which follow the formal assignment overseen by 
the IAU Minor Planet Center. 
HMAG and GPAR are the absolute magnitude and slope parameter taken
from the Asteroid Orbital Elements Database of the Lowell observatory. 
NID gives the number of detections at $S9W$ and $L18W$ in total.
DIAMETER and ALBEDO are the estimated size (diameter) and albedo, while
D\_ERR, A\_ERR are their uncertainties estimated from
the thermal model calculations.
Users should note objects with single detection (NID=1). 
Example of the catalog data is shown in Table \ref{table:catalog example}. 

\begin{table*}
\caption{Format of the AKARI Asteroid catalog}
\label{table:catalog contents}
\begin{center}
\begin{tabular}{rllll}
\hline
Column   & Format & Units    & Label     & Description\\\hline
  1 -  6 &  A6    & ---      & NUMBER   &  Asteroid's number\\
  8 - 25 &  A18   & ---      & NAME     &  Asteroid's name\\
 27 - 36 &  A10   & ---      & PROV\_DES&  Asteroid's provisional designation\\
 38 - 42 &  F5.2  & mag      & HMAG\footnotemark[\dag]     &  Absolute magnitude\\
 44 - 48 &  F5.2  & ---      & GPAR\footnotemark[\dag]     &  Slope parameter\\
 50 - 51 &  I2    & ---      & NID      &  Number of detections by AKARI\\
 53 - 59 &  F7.2  &  km      & DIAMETER &  Mean diameter\\
 61 - 65 &  F5.2  &  km      & D\_ERR   &  Uncertainty in diameter\\
 67 - 71 &  F5.3  & ---      & ALBEDO   &  Mean geometric albedo\\
 73 - 77 &  F5.3  & ---      & A\_ERR   &  Uncertainty in albedo\\
\hline
\multicolumn{5}{l}{\hbox{\parbox{100mm}{\footnotesize
      \par\noindent
      \footnotemark[\dag] The H-G values are taken from the Asteroid Orbital Elements Database 
      of the Lowell observatory. 
}\hss}}
\end{tabular}
\end{center}
\end{table*}

\begin{table*}
\caption{Example for the AcuA catalog data. Top 10 and bottom 10 asteroids in order of 
number and provisional designation of asteroids are listed.}
\label{table:catalog example}
\begin{center}
\begin{tabular}{rllrlrrrll}
\hline
{\footnotesize NUMBER} &%
{\footnotesize NAME} &%
{\footnotesize PROV\_DES} &%
{\footnotesize HMAG} &%
{\footnotesize GPAR} &%
{\footnotesize NID} &%
{\footnotesize DIAMETER} &%
{\footnotesize D\_ERR} &%
{\footnotesize ALBEDO} &%
{\footnotesize A\_ERR}\\
&%
&%
&%
{\footnotesize [mag]} &%
&%
&%
{\footnotesize [km]} &%
{\footnotesize [km]} &%
&%
\\\hline
1      &Ceres   & \multicolumn{1}{c}{}                 &  3.34 &  0.12&  7&  973.89& 13.31& 0.087& 0.003\\
2      &Pallas  & \multicolumn{1}{c}{}                 &  4.13 &  0.11& 12&  512.59&  4.98& 0.150& 0.004\\
3      &Juno    & \multicolumn{1}{c}{}                 &  5.33 &  0.32&  8&  231.09&  2.60& 0.246& 0.007\\
4      &Vesta   & \multicolumn{1}{c}{}                 &  3.20 &  0.32&  5&  521.74&  7.50& 0.342& 0.013\\
5      &Astraea & \multicolumn{1}{c}{}                 &  6.85 &  0.15&  7&  110.77&  1.37& 0.263& 0.008\\
6      &Hebe    & \multicolumn{1}{c}{}                 &  5.71 &  0.24& 11&  197.15&  1.83& 0.238& 0.006\\
7      &Iris    & \multicolumn{1}{c}{}                 &  5.51 &  0.15&  7&  254.20&  3.27& 0.179& 0.006\\
8      &Flora   & \multicolumn{1}{c}{}                 &  6.49 &  0.28& 10&  138.31&  1.37& 0.235& 0.006\\
9      &Metis   & \multicolumn{1}{c}{}                 &  6.28 &  0.17&  7&  166.48&  2.08& 0.213& 0.007\\
10     &Hygiea  & \multicolumn{1}{c}{}                 &  5.43 &  0.15&  6&  428.46&  6.57& 0.066& 0.002\\
\multicolumn{1}{r}{$\vdots$} &%
\multicolumn{1}{c}{$\vdots$} &%
\multicolumn{1}{c}{$\vdots$} &%
\multicolumn{1}{r}{$\vdots$\hspace{10pt} } &%
\multicolumn{1}{l}{\hspace{10pt}$\vdots$} &%
\multicolumn{1}{r}{$\vdots$\hspace{1.5pt} } &%
\multicolumn{1}{r}{$\vdots$\hspace{10pt} } &%
\multicolumn{1}{r}{$\vdots$\hspace{10pt} } &%
\multicolumn{1}{l}{\hspace{10pt}$\vdots$} &%
\multicolumn{1}{l}{\hspace{10pt}$\vdots$} \\
       &  \multicolumn{1}{c}{}       &           2006 SE285& 16.43 & 0.15&  1&    3.56&  0.30& 0.037& 0.006\\
       &  \multicolumn{1}{c}{}       &           2006 UD185& 14.39 & 0.15&  3&    8.76&  0.42& 0.048& 0.005\\
       &  \multicolumn{1}{c}{}       &           2006 UL217& 20.72 & 0.15&  1&    0.14&  0.01& 0.487& 0.073\\
       &  \multicolumn{1}{c}{}       &           2006 VV2  & 16.79 & 0.15&  1&    1.03&  0.03& 0.318& 0.024\\
       &  \multicolumn{1}{c}{}       &           2006 WT1  & 19.99 & 0.15&  1&    0.35&  0.02& 0.150& 0.018\\
       &  \multicolumn{1}{c}{}       &           2007 AG   & 20.11 & 0.15&  6&    0.33&  0.01& 0.158& 0.008\\
       &  \multicolumn{1}{c}{}       &           2007 BT2  & 17.06 & 0.15&  3&    2.76&  0.14& 0.038& 0.004\\
       &  \multicolumn{1}{c}{}       &           2007 DF8  & 20.32 & 0.15&  2&    0.47&  0.02& 0.059& 0.006\\
       &  \multicolumn{1}{c}{}       &           2007 FM3  & 16.87 & 0.15&  5&    3.14&  0.13& 0.033& 0.003\\
       &  \multicolumn{1}{c}{}       &           2007 HE15 & 19.60 & 0.15&  1&    0.37&  0.02& 0.182& 0.021\\
\hline
\end{tabular}
\end{center}
\end{table*}


\section{List of 55 asteroids used for thermal model calibration}
\label{List of 55 asteroids of calibrators}
We employ 55 well-studied main-belt asteroids (\cite{mueller05}) to
derive the best value for the beaming parameter $\eta$
(Sect.\ref{Thermal model calculation}).
Table \ref{table:55 asteroids} summarizes the calculation results of the 55 asteroids.


{\scriptsize
\begin{longtable}{rl|c||rrrrr|rrl}
\caption{Results of the STM calculation for the 55 selected asteroids. 
The references of previous work are given in Appendix \ref{Reference list}. 
The cited data refer to the underlined reference in the list.
}
\label{table:55 asteroids}
\hline
 & & & \multicolumn{5}{c|}{detection with AKARI} & & &\\
 & & & %
\multicolumn{1}{c}{$N_{\rm ID}$} & %
\multicolumn{1}{c}{$N_{\rm ID}$} & %
\multicolumn{1}{c}{$N_{\rm ID}$} & %
\multicolumn{2}{c|}{} &
\multicolumn{3}{c}{ previous work}  \\[-0pt]
\multicolumn{2}{c|}{Asteroid} & %
type & %
\multicolumn{1}{c}{$S9W$}& %
\multicolumn{1}{c}{$L18W$}& %
\multicolumn{1}{c}{total}& %
$d$[km] & $ p_{\rm v}$ & %
$d$[km] & $ p_{\rm v}$ & %
References \\\hline
\endhead
\hline
\endfoot
  1 &Ceres        & C & 3 & 4 & 7 &  973.89&0.087&  959.60&0.096 & { A1, D2, D4, D5, D7, }\\[-0pt]
    &             &   &   &   &   &        &     &        &      & { D8, D10, D16, D20, D26, }\\[-0pt]
    &             &   &   &   &   &        &     &        &      & { D29, D33, D34, D42, D52, }\\[-0pt]
    &             &   &   &   &   &        &     &        &      & { \underline{D67} }\\
  2 &Pallas       & C & 6 & 6 &12 &  512.59&0.150&  534.40&0.142 & { A1, D2, D5, D7, D15, }\\[-0pt]
    &             &   &   &   &   &        &     &        &      & { D16, D20, D22, D26, D33, }\\[-0pt]
    &             &   &   &   &   &        &     &        &      & { D42, D52, \underline{D67}}\\
  3 &Juno         & S & 4 & 4 & 8 &  231.09&0.246&  233.92&0.238 & { \underline{A1}, D2, D5, D26, D33, }\\[-0pt]
    &             &   &   &   &   &        &     &        &      & { D42, {D52}}\\
  4 &Vesta        & V & 2 & 3 & 5 &  521.74&0.342&  548.50&0.317 & { A1, D1, D2, D3, D4, }\\[-0pt]
    &             &   &   &   &   &        &     &        &      & { D5, D7, D8, D22, D25, }\\[-0pt]
    &             &   &   &   &   &        &     &        &      & { D26, D33, D34, D42, D52, }\\[-0pt]
    &             &   &   &   &   &        &     &        &      & { \underline{D67}}\\
  6 &Hebe         & S & 6 & 5 &11 &  197.15&0.238&  185.18&0.268 & { \underline{A1}, D2, D16, D26, {D34}}\\
  7 &Iris         & S & 3 & 4 & 7 &  254.20&0.179&  199.83&0.277 & { \underline{A1}, D3, D5, D15, D16, }\\[-0pt]
    &             &   &   &   &   &        &     &        &      & { D26, D34, {D42}}\\
  8 &Flora        & S & 4 & 6 &10 &  138.31&0.235&  135.89&0.243 & { \underline{A1}, D3, D5, {D26}}\\
  9 &Metis        & D & 4 & 3 & 7 &  166.48&0.213&  154.67&0.228 & { \underline{B1}, D3, D5, D42, D52, }\\[-0pt]
    &             &   &   &   &   &        &     &        &      & { {D55}}\\
 10 &Hygiea       & C & 3 & 3 & 6 &  428.46&0.066&  469.30&0.056 & { A1, D3, D5, D16, D18, }\\[-0pt]
    &             &   &   &   &   &        &     &        &      & { D22, D26, D33, D52, \underline{D67}}\\
 12 &Victoria     & S & 3 & 2 & 5 &  131.51&0.130&  112.77&0.176 & { \underline{A1}, D5, {D7}}\\
 17 &Thetis       & S & 4 & 2 & 6 &   74.59&0.251&   90.04&0.172 & { \underline{A1}, D3, {D5}}\\
 18 &Melpomene    & S & 3 & 3 & 6 &  139.95&0.225&  140.57&0.223 & { \underline{A1}, D3, D5, D34, {D52}}\\
 19 &Fortuna      & C & 3 & 3 & 6 &  199.66&0.063&  201.70&0.064 & { D3, D5, D7, D16, \underline{D55}}\\
 20 &Massalia     & S & 6 & 6 &12 &  131.56&0.258&  145.50&0.210 & { \underline{A1}, D5, D7, {D52}}\\
 21 &Lutetia      & X & 4 & 4 & 8 &  108.38&0.181&   95.76&0.221 & { \underline{A1}, D3, D5, D7, D59, }\\[-0pt]
    &             &   &   &   &   &        &     &        &      & { {D63} }\\
 23 &Thalia       & S & 1 & 3 & 4 &  106.21&0.260&  107.53&0.254 & { \underline{A1}, B1, D3, {D5}}\\
 24 &Themis       & C & 4 & 4 & 8 &  176.81&0.084&  176.20&0.084 & { D52, \underline{D55} }\\
 28 &Bellona      & S & 4 & 1 & 5 &   97.40&0.273&  120.90&0.176 & { \underline{A1}, B1, {D5}}\\
 29 &Amphitrite   & S & 3 & 4 & 7 &  206.86&0.195&  212.22&0.179 & { \underline{A1}, D3, D5, {D26}}\\
 31 &Euphrosyne   & C & 6 & 6 &12 &  276.49&0.047&  255.90&0.054 & { \underline{A1}}\\
 37 &Fides        & S & 3 & 3 & 6 &  103.23&0.204&  108.35&0.183 & { \underline{A1}, D3, {D5} }\\
 40 &Harmonia     & S & 3 & 5 & 8 &  110.30&0.233&  107.62&0.242 & { \underline{A1}, D3, D5, {D52}}\\
 41 &Daphne       & C & 3 & 4 & 7 &  179.61&0.078&  174.00&0.083 & { \underline{A1}, {D7}}\\
 42 &Isis         & S & 4 & 3 & 7 &  104.50&0.158&  100.20&0.171 & { \underline{A1}, {D7}}\\
 47 &Aglaja       & C & 2 & 1 & 3 &  147.05&0.060&  126.96&0.080 & { \underline{A1}, {D5}}\\
 48 &Doris        & C & 3 & 4 & 7 &  200.27&0.077&  221.80&0.062 & { \underline{A1}}\\
 52 &Europa       & C & 4 & 3 & 7 &  350.36&0.043&  302.50&0.058 & { \underline{A1}, D5, D7, {D26}}\\
 54 &Alexandra    & C & 3 & 5 & 8 &  144.46&0.074&  165.75&0.056 & { \underline{A1}, D5, D7, D33, {D52}}\\
 56 &Melete       & X & 4 & 6 &10 &  105.22&0.076&  113.24&0.065 & { \underline{A1}, D5, {D16}}\\
 65 &Cybele       & X & 4 & 2 & 6 &  300.54&0.044&  237.26&0.071 & { \underline{A1}, D16, D26, {D33}}\\
 69 &Hesperia     & X & 5 & 4 & 9 &  132.74&0.157&  138.13&0.140 & { \underline{A1}}\\
 85 &Io           & C & 4 & 4 & 8 &  150.66&0.071&  154.79&0.067 & { \underline{A1}, {D7}}\\
 88 &Thisbe       & C & 3 & 4 & 7 &  195.59&0.071&  200.58&0.067 & { \underline{A1}}\\
 93 &Minerva      & C & 3 & 3 & 6 &  147.10&0.068&  141.55&0.073 & { \underline{A1}, {B1}}\\
 94 &Aurora       & C & 2 & 2 & 4 &  179.15&0.053&  204.89&0.040 & { \underline{A1}, D5, {D7}}\\
106 &Dione        & C & 3 & 3 & 6 &  153.42&0.084&  146.59&0.089 & { \underline{A1}, D7, {D33}}\\
165 &Loreley      & C & 4 & 2 & 6 &  173.66&0.051&  154.78&0.064 & { \underline{A1}}\\
173 &Ino          & X & 3 & 1 & 4 &  160.61&0.059&  154.10&0.064 & { \underline{A1}}\\
196 &Philomela    & S & 2 & 4 & 6 &  141.78&0.213&  136.39&0.230 & { \underline{A1}, D5, {D7}}\\
230 &Athamantis   & S & 4 & 5 & 9 &  108.28&0.173&  108.99&0.171 & { \underline{A1}, D3, D5, {D7}}\\
241 &Germania     & C & 3 & 3 & 6 &  181.57&0.050&  168.90&0.058 & { \underline{A1}, {D5}}\\
283 &Emma         & C & 4 & 8 &12 &  122.07&0.039&  148.06&0.026 & { \underline{A1}}\\
313 &Chaldaea     & C & 4 & 4 & 8 &   94.93&0.054&   96.34&0.052 & { \underline{A1}, D5, D8, {D33}}\\
334 &Chicago      & C & 4 & 5 & 9 &  167.21&0.057&  158.55&0.062 & { \underline{A1}}\\
360 &Carlova      & C & 4 & 4 & 8 &  121.52&0.049&  115.76&0.053 & { \underline{A1}, D5, {D7}}\\
372 &Palma        & C & 2 & 4 & 6 &  177.21&0.075&  188.62&0.066 & { \underline{A1}}\\
423 &Diotima      & C & 5 & 1 & 6 &  226.91&0.049&  208.77&0.051 & { \underline{A1}}\\
451 &Patientia    & C & 5 & 5 &10 &  234.91&0.071&  224.96&0.076 & { \underline{A1}, D5, D7, {D16}}\\
471 &Papagena     & S & 3 & 3 & 6 &  117.44&0.261&  134.19&0.199 & { \underline{A1}, {D3}}\\
505 &Cava         & C & 5 & 4 & 9 &  100.55&0.063&  115.80&0.040 & { \underline{D55}}\\
511 &Davida       & C & 4 & 3 & 7 &  290.98&0.070&  326.06&0.054 & { \underline{A1}, D2, D3, D7, {D52}}\\
532 &Herculina    & S & 4 & 2 & 6 &  216.77&0.184&  222.39&0.169 & { \underline{A1}, D3, D5, D8, D33, }\\[-0pt]
    &             &   &   &   &   &        &     &        &      & { {D42}}\\
690 &Wratislavia  & C & 2 & 4 & 6 &  158.11&0.044&  134.65&0.060 & { \underline{A1}}\\
704 &Interamnia   & C & 7 & 4 &11 &  316.25&0.075&  316.62&0.074 & { \underline{A1}, D5, D7,{D52}}\\
776 &Berbericia   & C & 4 & 5 & 9 &  149.76&0.067&  151.17&0.066 & { \underline{A1}}\\
\end{longtable}
}


\newpage

\section{Reference list of previous work of the size and albedo of asteroids}
\label{Reference list}
The references of previous work are given in the following list.\\
%
The Infrared Astronomy Satellite (IRAS):\\
{\scriptsize
(A1) ~\cite{SIMPS}.\\
}

The Midcourse Space Experiment (MSX)\\
{\scriptsize
(B1) ~\cite{MIMPS}.\\
}

The Spitzer Space Telescope (SST):\\
{\scriptsize
  \begin{tabular}{ll}
    (C1) ~\cite{St2008}&  (C2) ~\cite{Tr2008} \\
    (C3) ~\cite{Ry2009}&  (C4) ~\cite{Ca2009a} \\
    (C5) ~\cite{Ca2009b}& (C6) ~\cite{Fe2009} \\
    (C7) ~\cite{Ha2009}&  (C8) ~\cite{Li2009} \\
    (C9) ~\cite{Bh2010}&  (C10) ~\cite{Tr2010}\\
  \end{tabular}\\
}

Other work including the Infrared Space Observatory (ISO) and 
ground-based observatories in the chronological order:

$*$ 1970--1979\\
{\scriptsize
  \begin{tabular}{ll}
    (D1) ~\cite{Al1970} & (D2) ~\cite{Cr1973}\\
    (D3) ~\cite{Mo1974} & (D4) ~\cite{Gi1975}\\
    (D5) ~\cite{Ha1976} & (D6) ~\cite{Cr1977}\\
    (D7) ~\cite{Mo1977} & (D8) ~\cite{Gr1978}\\
    (D9) ~\cite{Le1978} & (D10) ~\cite{St1978}\\
    (D11) ~\cite{Le1979}\\
  \end{tabular}\\
}

$*$ 1980--1989\\
{\scriptsize
  \begin{tabular}{ll}
    (D12) ~\cite{Le1981}  &(D13) ~\cite{Br1984}\\
    (D14) ~\cite{Le1984}  &(D15) ~\cite{LV1984}\\
    (D16) ~\cite{Gr1985a} &(D17) ~\cite{Gr1985b}\\
    (D18) ~\cite{Le1985}  &(D19) ~\cite{Vi1985}\\
    (D20) ~\cite{lebofsky86}  &(D21) ~\cite{Te1987}\\
    (D22) ~\cite{Jo1989}  &(D23) ~\cite{Ve1989}\\
  \end{tabular}\\
}

$*$ 1990--1999\\
{\scriptsize
  \begin{tabular}{ll}
    (D24) ~\cite{Cr1991}& (D25) ~\cite{Re1992}\\
    (D26) ~\cite{Al1994}& (D27) ~\cite{Al1995}\\
    (D28) ~\cite{Ca1995}& (D29) ~\cite{Al1996}\\
    (D30) ~\cite{Mo1997}& (D31) ~\cite{Ha1998}\\
    (D32) ~\cite{Je1998}& (D33) ~\cite{Mu1998}\\
    (D34) ~\cite{Re1998}& (D35) ~\cite{Ha1999}\\
  \end{tabular}\\
}

$*$ 2000--2009\\
{\scriptsize
  \begin{tabular}{ll}
    (D36) ~\cite{Th2000} & (D37) ~\cite{Al2001}\\
    (D38) ~\cite{Fe2001} & (D39) ~\cite{Ha2001}\\
    (D40) ~\cite{Je2001} & (D41) ~\cite{Fe2002}\\
    (D42) ~\cite{Mu2002} & (D43) ~\cite{Te2002}\\
    (D44) ~\cite{De2002} & (D45) ~\cite{Fe2003}\\
    (D46) ~\cite{De2004} & (D47) ~\cite{Mu2004}\\
    (D48) ~\cite{Cr2005} & (D49) ~\cite{Fe2005}\\
    (D50) ~\cite{Ha2005} & (D51) ~\cite{Kr2005}\\
    (D52) ~\cite{Li2005} & (D53) ~\cite{Mu2005}\\
    (D54) ~\cite{Ri2005} & (D55) ~\cite{Te2005}\\
    (D56) ~\cite{WO2005} & (D57) ~\cite{De2006}\\
    (D58) ~\cite{Em2006} & (D59) ~\cite{Mu2006}\\
    (D60) ~\cite{Ha2007} & (D61) ~\cite{Mu2007}\\
    (D62) ~\cite{Tr2007} & (D63) ~\cite{Ca2008}\\
    (D64) ~\cite{Ha2008} & (D65) ~\cite{WO2008}\\
    (D66) ~\cite{De2009} & (D67) ~\cite{Ho2009}\\
  \end{tabular}\\
}

\section{Reference list of previous work of the taxonomic types of asteroids}
\label{Reference list for taxonomic types}
The references of previous work used for determining the taxonomic classifications 
(based on the definitions by \citet{tholen84}, \citet{bus99}, and \citet{lazzaro04}) 
described in Table \ref{table:summary for 5 types} are given in the following list.\\

{\scriptsize
  \begin{tabular}{ll}
    (E1)  ~\cite{Jewitt90}        & (E2)  ~\cite{Barucci93}\\
    (E3)  ~\cite{dahlgren95}      & (E4)  ~\cite{Xu95}\\
    (E5)  ~\cite{Dahlgren97}      & (E6)  ~\cite{Martino97}\\
    (E7)  ~\cite{Lazzarin97}      & (E8)  ~\cite{Doressoundiram98}\\
    (E9)  ~\cite{Hicks98}         & (E10) ~\cite{Hicks00}\\
    (E11) ~\cite{Zappala00}       & (E12) ~\cite{binzel01}\\
    (E13) ~\cite{Cellino01}       & (E14) ~\cite{Fornasier01}\\
    (E15) ~\cite{Le Bras01}       & (E16) ~\cite{Manara01}\\
    (E17) ~\cite{Mothe-Diniz01}   & (E18) ~\cite{Fornasier03}\\
    (E19) ~\cite{rivkin03}        & (E20) ~\cite{Yang03}\\
    (E21) ~\cite{Bendjoya04}      & (E22) ~\cite{binzel04}\\
    (E23) ~\cite{Duffard04}       & (E24) ~\cite{Fornasier04}\\
    (E25) ~\cite{Marchi04}        & (E26) ~\cite{lazzarin04}\\
    (E27) ~\cite{lazzarin04b}     & (E28) ~\cite{Lagerkvist05}\\
    (E29) ~\cite{lazzarin05}      & (E30) ~\cite{marchi05}\\
    (E31) ~\cite{Alvarez06}       & (E32) ~\cite{Dotto06}\\
    (E33) ~\cite{Leon06}          & (E34) ~\cite{Michelsen05}\\
    (E35) ~\cite{binzel07}        & (E36) ~\cite{Licandro08} \\
    (E37) ~\cite{moskovitz08}     & (E38) ~\cite{moskovitz08b}\\
    (E39) ~\cite{monthe-diniz08}  & (E40) ~\cite{monthe-diniz08b}\\
    (E41) ~\cite{roig08}          & (E42) ~\cite{duffard09}\\
  \end{tabular}\\
}


\end{document}